\documentclass[%
 reprint,
superscriptaddress,
 amsmath,amssymb,
 aps,
]{revtex4-2}

\usepackage[utf8]{inputenc}
\usepackage{lmodern} 
\usepackage{microtype}
\usepackage{natbib}
\setcitestyle{numbers,square,sort&compress}
\usepackage{amsmath,amssymb,amsfonts}
\usepackage{algorithmic}
\usepackage{graphicx}
\usepackage{textcomp}
\usepackage{lipsum,mathtools,cuted}
\usepackage{xcolor}
\usepackage{hyperref}

\usepackage{float}
\usepackage{epsfig}
\usepackage{epstopdf}
\usepackage{subfigure,color}
\usepackage{dcolumn}
\usepackage{bm}
\usepackage{lineno}
\usepackage{accents}
\usepackage{multirow}
\usepackage[usestackEOL]{stackengine}
\usepackage{tabularx}
\usepackage{makecell}
\usepackage{dcolumn}% Align table columns on decimal point
\usepackage{bm}% bold math
\usepackage{booktabs}

\UseRawInputEncoding

\usepackage{mdframed}

\newcommand*{\colorboxed}{}
\def\colorboxed#1#{%
  \colorboxedAux{#1}%
}
\newcommand*{\colorboxedAux}[3]{%
  \begingroup
    \colorlet{cb@saved}{.}%
    \color#1{#2}%
    \boxed{%
      \color{cb@saved}%
      #3%
    }%
  \endgroup
}

\begin{document}
\title{Extreme nonreciprocity  in metasurfaces based on bound states in the continuum}

\author{L. M. Ma\~{n}ez-Espina}
\affiliation{
Nanophotonics Technology Center, Universitat Polit\`ecnica de Val\`encia, Valencia 46022, Spain
}
\email{lmmaeesp@upv.es}
 \author{I. Faniayeu}
\affiliation{
Department of Physics, University of Gothenburg, Gothenburg 41296, Sweden
}
\author{V. Asadchy}
\affiliation{Department of Electronics and Nanoengineering, Aalto University, P.O.~Box 15500, FI-00076 Aalto, Espoo, Finland}
\author{A. D\'iaz-Rubio}
\affiliation{
Nanophotonics Technology Center, Universitat Polit\`ecnica de Val\`encia, Valencia 46022, Spain
}

\date{\today}% It is always \today, today,
             %  but any date may be explicitly specified

%\communicated{...}
%\dedication{...}
	
\begin{abstract}
% Developing   nonreciprocal   devices is  very important    for   optical systems such as   isolators, phase shifters and optical amplifiers. Unfortunately, magneto-optical effects in natural materials are very weak, leading to the necessity to have large-thickness nonreciprocal devices. In this work, we 
% propose and demonstrate that by using metasurfaces made out of  a realistic magnetic ferrite supporting bound states in the continuum, it is possible to achieve extreme nonreciprocity in the Faraday configuration and near-unity magnetic circular dichroism. The magneto-optical effect in the proposed metasurface exceeds that in a continuous film of the same material by 3–4 orders of magnitude. Such an enhancement is provided  by exploiting   Huygens’ condition  in the metasurface that supports a 
% pair of electric and magnetic  dipole resonances.   
% Our findings are supported by  a phenomenological
% multi-mode temporal coupled mode theory which is in full agreement with full-wave simulation results.
%
Nonreciprocal devices, including optical isolators, phase shifters, and amplifiers, are pivotal for advanced optical systems. However, exploiting natural materials is challenging due to their weak magneto-optical effects, requiring substantial thickness to construct effective optical devices. In this study, we demonstrate that subwavelength metasurfaces supporting bound states in the continuum and made of conventional magnetic ferrite can exhibit extreme nonreciprocity in the Faraday configuration and near-unity magnetic circular dichroism. These metasurfaces enhance the magneto-optical effect by 3--4 orders of magnitude compared to a continuous film of the same material. This significant enhancement is achieved by leveraging Huygens' condition in the metasurface whose structural units support paired electric and magnetic dipole resonances. We develop the multi-mode temporal coupled-mode theory for the observed enhancement of the magneto-optical effect and confirm our findings with the full-wave simulations.
% Developing nonreciprocal optical devices is crucial for the implementation of some of the basic building blocks of complex optical systems, such as isolators, circulators, and directional amplifiers. One of the most exploited ways to achieve nonreciprocal behavior is the use of magneto-optical materials. However, the magneto-optical effects that these materials present are weak in the optical domain, imposing devices to be bulky in comparison with their working wavelength. In recent years, the usage of metasurfaces has been investigated as a way to minimize the size of devices utilizing this physical phenomenon. 
% In our work, we propose and demonstrate that by using metasurfaces made out of Bi3YIG (a magnetic ferrite) supporting quasi-BICs, it is possible to achieve extreme nonreciprocity and huge enhancement of the Faraday effect.  When illuminated by circularly polarized light, this metasurface alone can also function as a one-way system, achieving strong circular dichroism and responding differently on each side. A phenomenological coupled mode theory model and finite element simulations with realistic permittivity values are used to support the demonstration.
\end{abstract}

\keywords{All-dielectric metasurfaces, non-local metasurfaces, coupled mode theory.}

\maketitle

\section{Introduction}

In the race for the miniaturization of optical devices, a wide class of electromagnetic structures named metasurfaces arose in the last decade as one of the most prominent solutions. Optical metasurfaces are two-dimensional arrays of nano-inclusions with a sub-wavelength thickness that are engineered to manipulate light at will~\cite{Roadmap-meta,Review-1,Review-2,Review-3}. Theoretically, if the inclusions, known as meta-atoms, are properly designed, almost any response that does not violate fundamental laws of physics could be achieved for these structures. One of the important industrial applications of metasurfaces is the design of nonreciprocal optical components such as isolators. In order to achieve optical isolation, a necessary condition is that the device must be nonreciprocal~\cite{Isolator}. Traditionally, nonreciprocity has been implemented using Magneto-Optical (MO) materials such as ferrites~\cite{Nonreciprocity}. Nevertheless, MO effects are very weak in the optical regime~\cite{MO-physics} and must be enhanced with proper techniques for designing compact nonreciprocal devices. Recently, there have been multiple techniques suggested to achieve an enhancement of MO effects at the nanoscale: magnetophotonic crystals~\cite{MO-magnetophotoniccrstals,MO-PHC-1,MO-PHC-2,MO-PHC-3}, coupled surface-plasmon polaritons~\cite{MO-magnetoplasmonics2,MO-Plasmonics,MO-SPPs}, magnetic Weyl semimetals~\cite{asadchy_sub_2020}, multiple layer systems~\cite{MO-layer}, and resonant all-dielectric metasurfaces~\cite{MO-Huygens-FOM, MO-meta-FOM,MO-hybrid,MO-BICs-3}. 
Although these studies achieve a high enhancement, the resonant properties of the structures considered do not allow ultra-high quality factors that would enable higher interaction times between light and the MO material~\cite{MO-meta-FOM}. 
%Although these studies achieve a high enhancement, their regular resonances do not allow ultra-high quality factors that would enable higher interaction times between light and the MO material~\cite{MO-meta-FOM}. 

%An attractive path toward designing compact optical nonreciprocal components is based on bound states in the continuum (BICs) which, if properly engineered, enable almost non-leaky resonances with high confinement inside the device's material. 
An attractive path toward designing compact optical nonreciprocal components is based on bound states in the continuum (BICs) which, if properly engineered, enable almost non-leaky resonances that increase the interaction time and the field concentration in the materials of the resonant structure.
BICs in metasurfaces have been widely researched in the last few years~\cite{Meta-BICs}. In electromagnetism, these bound states with infinite lifetimes are discrete solutions of the Maxwell equations embedded in a continuum of leaky modes and inaccessible by incident external radiation~\cite{BICs-review}. Their useful realization comes in the form of quasi-BICs, which are perturbed BICs that have turned into accessible leaky modes with extreme quality factors (ultra-long but finite lifetimes)~\cite{BIC-1,BIC-2}. Over recent years, metasurfaces that use these states have been proposed to enhance different effects: circular dichroism~\cite{BIC-CD,Maximun-chirality-kivshar,BICs-chirality}, absorption~\cite{BIC-absorption,BIC-absorption2}, and more. Furthermore, BICs can be classified into different categories depending on their formation nature~\cite{BICs-review}. Particularly interesting and relatively easy to exploit are the Symmetry-Protected BICs (SP-BICs)~\cite{BICs-SP}, which have been precisely characterized by their group representations~\cite{BICs-grupos,BICs-SP-2,BICs-SP-3}. Specifically, SP-BICs are inaccessible states that do not couple to the plane wave modes because of symmetry incompatibility between incident excitation and structural symmetries~\cite{BIC-2}. Several metasurface designs with enhanced MO effect were recently proposed based on BICs~\cite{MO-BICs-2,MO-BICs, abujetas_active_2021}. However, they were either limited to the THz range where magneto-optical effects are relatively strong even in natural materials~\cite{MO-BICs-2}, or based on nonrealistic materials (exhibiting no dissipation loss)~\cite{abujetas_active_2021}, or exhibited relatively weak MO properties (in~\cite{MO-BICs}, the reported Faraday rotation was in the order of 12~mrad). More importantly, in none of these works, a comprehensive theoretical explanation of the MO enhancement mechanism was given, which is essential for optimizing the performance of the structure and reaching the maximum strength of the effect.

%In this work, we start by developing a phenomenological multi-mode temporal coupled mode theory (CMT)~\cite{CMT-OG,CMT-eqs} to serve as a recipe for the design of optical metasurfaces exhibiting maximum MO effects for the given materials being used. We demonstrate that by exploiting a pair of even and odd resonances in the metasurface unit cell (so-called Huygens' pair~\cite{Huygens-pairs}), it is possible to obtain nearly unit magnetic circular dichroism (MCD) in a simple geometry with conventional material (${\rm Bi}_3{\rm YIG}$) being used. We design a nonreciprocal metasurface with a near-unit MCD that behaves as a one-way nonreciprocal device for circularly polarized light, allowing a certain handedness to be transmitted when propagating in the forward direction and blocking it for backward illumination. If surrounded by two polarizers~\cite{Geometry-1}, such a structure would operate as an optical isolator in transmission. 
In this work, we present a metasurface made of MO ferrite material that simultaneously supports BICs with electric and magnetic nature.  
We delve into the physical phenomena in the structure and develop a phenomenological multi-mode temporal coupled mode theory (CMT)~\cite{CMT-OG,CMT-eqs} to serve as a recipe for the design of optical metasurfaces exhibiting maximum MO effects for the given materials being used.
We demonstrate that by exploiting a pair of even and odd resonances in the metasurface unit cell (so-called Huygens' pair~\cite{Huygens-pairs}), it is possible to obtain nearly unit magnetic circular dichroism (MCD) in a simple geometry with conventional material (${\rm Bi}3{\rm YIG}$) being used.
Using this recipe, we design a nonreciprocal metasurface with a near-unit MCD that behaves as a one-way nonreciprocal device for circularly polarized light, allowing a certain handedness to be transmitted when propagating in the forward direction and blocking it for backward illumination. 
If surrounded by two polarizers~\cite{Geometry-1}, such a structure would operate as an optical isolator in transmission. 

%-----------------------------------------------------------------------------------

\section{Metasurface Design}

\begin{figure}[t!]
    \centering 
    \includegraphics[width=\columnwidth]{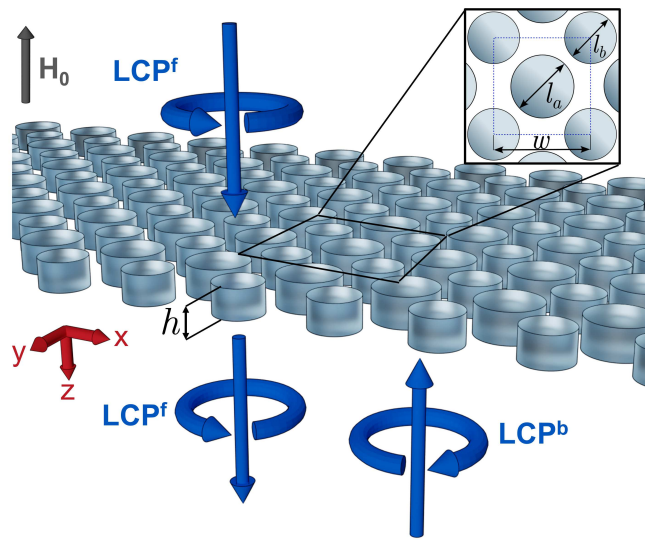}
    \caption{Schematic of the nonreciprocal metasurface exhibiting one-way transmission for circularly polarized light. The inset depicts the unit cell with the in-plane geometrical parameters. The neighboring nanodisks have slightly different diameters $l_a$ and $l_b$ and are arranged in checkerboard order. The LCP plane wave is labeled as ``f" (forward) and ``b" (backward). The external magnetic bias necessary to break reciprocity is shown at the top left of the figure.}
    \label{fig: Geometry}
\end{figure}

The metasurface design will follow a similar topology to that of~\cite{Geometry-1,Geometry-2}, where they described an array of dielectric nanodisks distributed in a squared lattice in a checkerboarded fashion with an asymmetry in diameter between nearest neighbors. However, in the present scenario, we consider the nanodisks to be made of MO ferrite material ${\rm Bi3YIG}$ experimentally characterized in ~\cite{Meta-BiYIG-material}. The metasurface geometry is shown in Fig.~\ref{fig: Geometry}. Its geometrical parameters are $l$ for the unperturbed diameter of the nanodisks, $w$ for the squared lattice constant, $h$ for the height of the nanodisks, and $\Delta$ for the difference in diameter between nearest neighbors. The diameters for the nanodisks are defined as $l_a=l+\Delta/2$ and $l_b=l-\Delta/2$. For simplicity of the analysis, we assume the metasurface to be surrounded by air from both sides. 
It is worth noticing that the same physics applies to the scenario when the metasurface is immersed inside some background dielectric material (for that, merely the permittivity of the nanodisks should be scaled up accordingly). For the scenario where the metasurface is located on top of a dielectric substrate, the theory can also be straightforwardly extended by considering non-resonant reflections from the substrate. 

The  material of the nanodisks is magnetized by a bias static magnetic field applied along the $+z$-axis, which results in its antisymmetric permittivity tensor of the form~\cite{Nonreciprocity}
\begin{equation}\label{eq:permittivity-tensor}
    \bar{\bar\varepsilon}(\omega,\textbf{H}_0 )=
    \begin{pmatrix}
    \varepsilon_{xx}(\omega) & j\varepsilon_{xy}(\omega,\textbf{H}_0 ) & 0\\
    -j\varepsilon_{xy}(\omega,\textbf{H}_0 )\ & \varepsilon_{xx}(\omega) & 0\\
    0 & 0 & \varepsilon_{zz}(\omega)
    \end{pmatrix},
\end{equation}
where $\varepsilon_{xx}=\varepsilon_{1r}-j\varepsilon_{1i}$, and $\varepsilon_{xy}=\varepsilon_{2r}-j\varepsilon_{2i}$ are complex permittivities and $\textbf{H}_0$ denotes an external magnetic field along the $z$-axis. Here we use the $e^{+j \omega t}$ convention for temporal field evolution. Permittivity component $\varepsilon_{xy}$ is assumed to be a linear function of $\textbf{H}_0$. Moreover, saturation is achieved in Bi3YIG with a field of 1.2 T~\cite{Meta-BiYIG-material}.

\begin{figure}[t!]
    \centering
    \subfigure[]{\includegraphics[width=0.45\columnwidth]{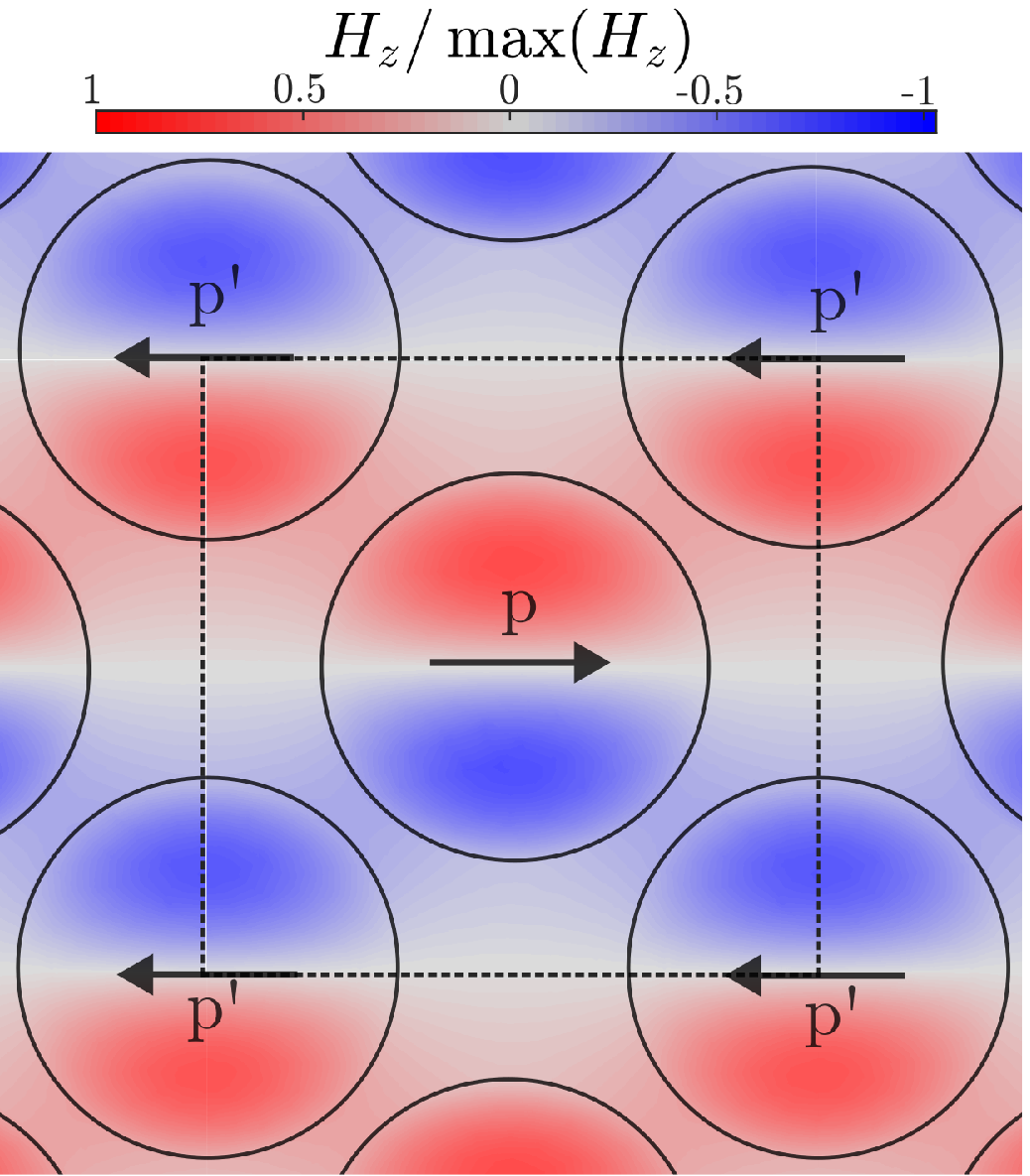}} 
    \subfigure[]{\includegraphics[width=0.45\columnwidth]{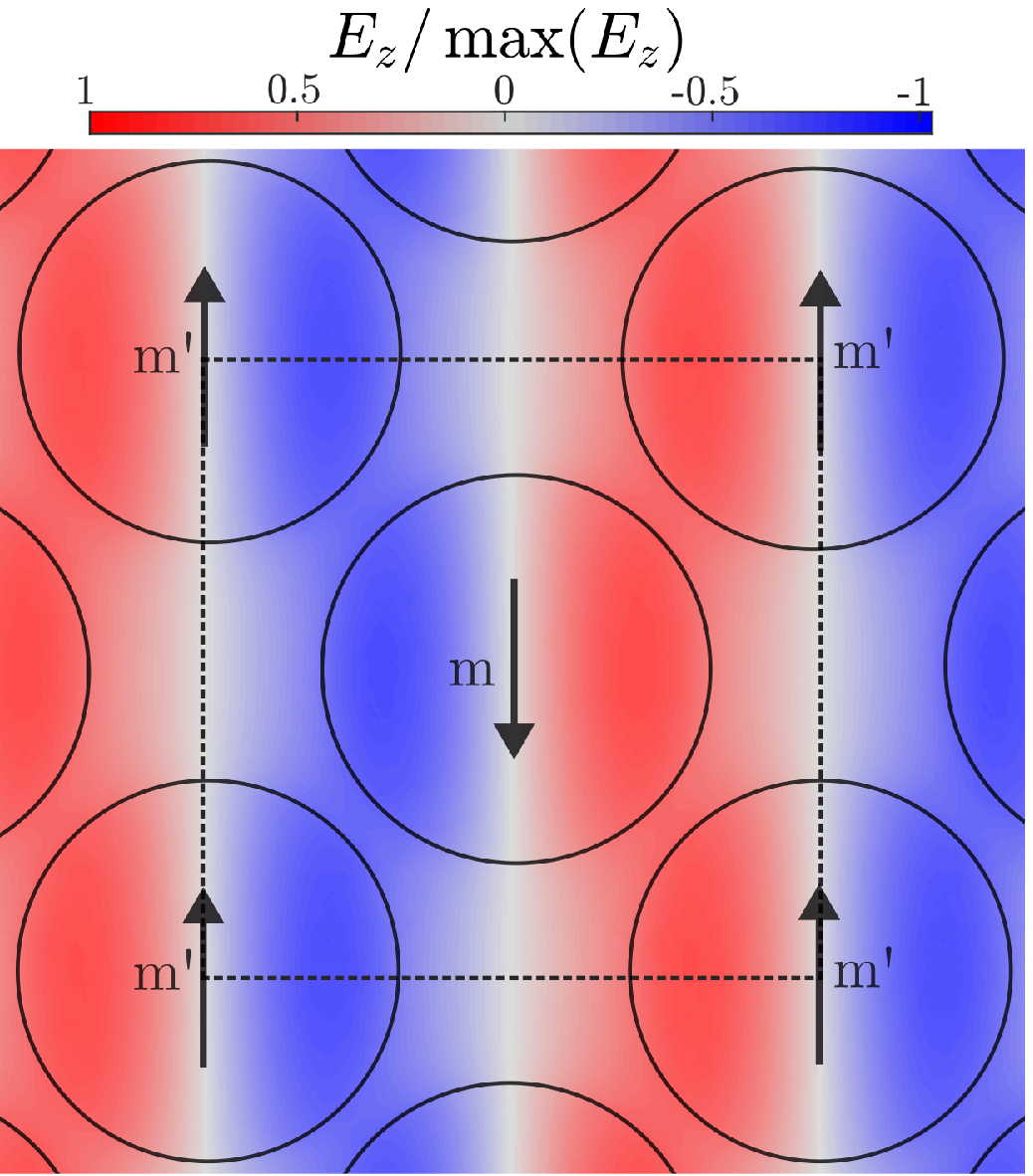}}
    \subfigure[]{\includegraphics[width=0.9\columnwidth]{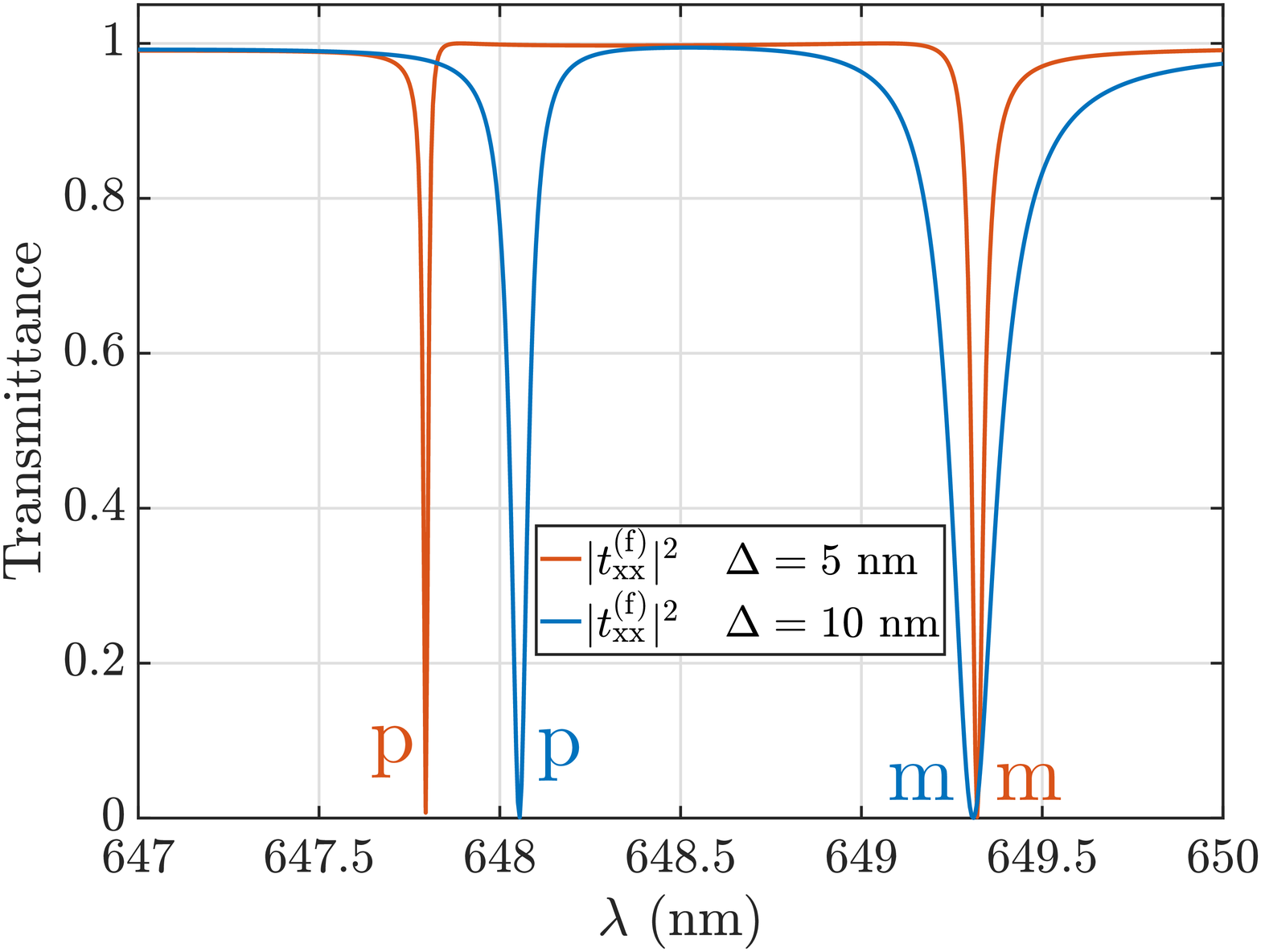}} 
    \caption{Resonant properties of the metasurface without external magnetic bias. Total field supported by the structure when $\Delta=5,10$ nm, $l=270$ nm, $w=430$ nm, $h=135.2$ nm, $\varepsilon_{xy}=0$, and $\varepsilon_{xx}=\varepsilon_{1r}$ in the plane $z=0$ (cutting-plane in the middle of the nanodisks).  (a) Electric dipole mode at $\lambda=647.73$~nm and (b) magnetic dipole mode at $\lambda=649.30$~nm in the case where $\Delta=5$~nm. The lattice cell is depicted with a dashed line. The dipole direction is shown with arrows. 
    (c) Transmittance of the metasurface without magnetic bias and dissipation losses. The asymmetry in diameter is proportional to the width of the resonance. The magnetic resonance width is greater than the electric dipole resonance. There is a noticeable shift in the electric dipole resonance frequency when $\Delta$ is changed.}
    \label{fig:BICs_Delta_Nobias_noloss}
\end{figure}

When $\varepsilon_{xy}=0$ and $\varepsilon_{xx}=\varepsilon_{1r}$, the structure shown in Fig.~\ref{fig: Geometry} is known to hold quasi-BICs resonant modes at normal incidence due to the mismatch in diameter between nanodisks~\cite{Geometry-1,Geometry-2}. The resonant modes of the metasurface that are those shown in Fig.~\ref{fig:BICs_Delta_Nobias_noloss}(a) and (b), obtained using the commercial FEM simulator COMSOL Multiphysics$^\circledR$. These two modes correspond to the scenarios where the nanodisks acquire electric and magnetic dipole polarizations, respectively. These modes remain inaccessible for normal plane-wave incidence in the metasurface without geometrical perturbations (i.e., $\Delta=0$). They are symmetry-protected bound states in the continuum (SP-BICs)~\cite{BICs-SP}, and their origin is due to the symmetries of the structure and can be explained in several ways. The most intuitive explanation can be done in terms of induced dipoles. A checkerboarded array of nanodisks can be thought of as a superposition of two sub-lattices, in this case, each holding a magnetic/electric dipole moment, but with a $\pi$ relative phase between wave oscillations in the two lattices. In the unperturbed case, the superposition of both lattices cancels the far-field radiation and forms an SP-BIC. In the perturbed case, the size of the nearest neighbor is altered, and the electric/magnetic dipoles are not identical. This geometrical mismatch induces a difference in amplitude and/or phase, enabling a non-perfect destructive superposition of scattered waves and creating a high-Q resonance, i.e., a quasi-BIC. When the perturbed metasurface supporting this type of resonant mode is illuminated by a linearly polarized wave,  the interaction of the incident field and the resonant structure will produce two peaks in the transmission spectra, as it is shown in Fig.~\ref{fig:BICs_Delta_Nobias_noloss}(c). These two peaks correspond to the resonance conditions of the two sub-lattices of dipole moments. The positions and the widths of the transmission peaks are closely related to the asymmetry of the structure quantified by the parameter $\Delta$. The resonant mode that appears at a lower wavelength is the electric dipole. As expected, the width of the resonant modes increases with the parameter $\Delta$ as the structure becomes more asymmetric. It is also interesting to notice that because the parameter $\Delta$ is directly related to the radius of the nanodisks, it has a larger impact on the position of the electric dipole. 

In a more formal mathematical fashion, the BIC formation phenomenon can be explained by using group theory techniques. In the case of $\Delta=0$, $\varepsilon_{xx}=\varepsilon_{1r}$, and $\varepsilon_{xy}=0$, the cell and point symmetries are $C_{4v}$ and $C_2$ with two additional mirror planes (see Supplementary Material~\cite{Supp}, Section 2). As it is known, the eigenmodes of a periodic structure must satisfy the transformations given by an \textit{irreducible representation} (IRREP) of the structure's symmetry group~\cite{BICs-SP-2}. There is a BIC when there is no coupling between the external excitation and the structure's eigenmode. The absence of coupling is due to the incompatibility between the symmetry restrictions of the resonant structure and the incoming plane wave's symmetries~\cite{Group-Theory-Phc}. Introducing a geometrical perturbation (i.e., $\Delta \neq 0$) removes the $C_2$ symmetry and one of the mirror planes, Fig. 3 in Supplementary material~\cite{Supp}. Without them, the eigenmode symmetry is partially broken, and the coupling is enabled. The state is now referred to as a quasi-BIC. For this reason, the symmetry-protected label we used to refer to these BICs is now apparent. The structure behaves exactly the same way with linear $y$-polarized light or $x$-polarized light due to the $C_{4v}$ symmetry that it still holds. 
%Moreover, the eigenmodes   shown in Fig.~\ref{fig:BICs_Delta_Nobias_noloss}(a) and (b) transform according to the symmetric \red{A} irreducible representation of the $C_2$ group \cite{BICs-chirality}. This symmetry emerges as a pair of antiparallel dipole moments. 
Moreover, the cell symmetry $C_{4v}$ explains the degeneracy of the electric and magnetic modes that belong to the same $\Gamma_5$ two-dimensional IRREP~\cite{BICs-grupos} (see Supplementary Materials~\cite{Supp}, Section 2). Furthermore, these eigenmodes, shown in Fig.~\ref{fig:BICs_Delta_Nobias_noloss}(a) and (b), transform according to the $\Gamma_5$ irreducible representation of the $C_{4v}$ group~\cite{BICs-chirality}.

\begin{figure}[t]
    \centering
\includegraphics[width=1\columnwidth]{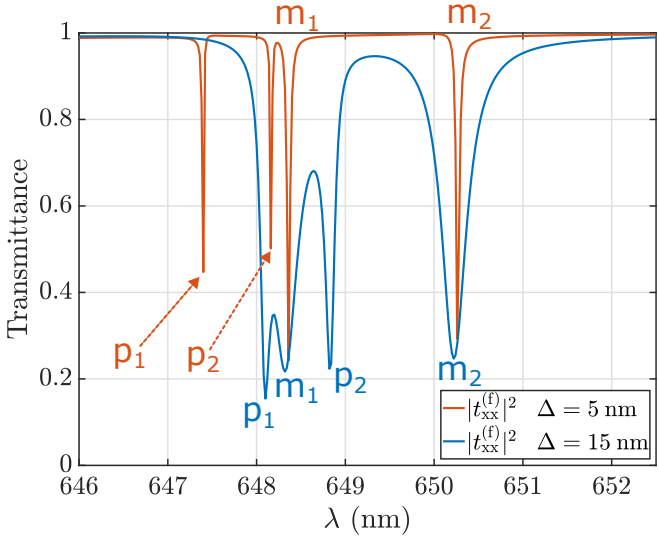}
\caption{Resonant properties of the metasurface with external magnetic bias but without dissipation losses.  The transmittance of the magnetically biased perturbed metasurface when {$\Delta=5 \rm nm$} and {$\Delta=15 \rm nm$} with the rest of parameters with values $l=270$ nm, $w=430$ nm and $h=135.2$ nm,. The electric dipoles $\rm  p_1$ and $\rm p_2$ suffer a higher shift in wavelength than the magnetic dipole resonances $\rm m_1$ and $\rm m_2$.}
    \label{fig:Linear-biased}
\end{figure}
 With the introduction of a magnetic bias, but neglecting the losses ($\varepsilon_{xy}=\varepsilon_{2r}$ and $\varepsilon_{xx}=\varepsilon_{1r}$), both magnetic and electric dipolar resonances split. As an example of this behavior, Fig.~\ref{fig:Linear-biased} represents the transmission spectra for two different values of the parameter $\Delta$ when the structure is biased by a magnetic field that brings the magnetic ferrite to saturation. Note that this analysis is done neglecting the losses in the material. The degeneracy that emerged from the two-dimensional $\Gamma_5$ IRREP is now broken by the non-equal off-diagonal components of the permittivity tensor (see Eq.~\ref{eq:permittivity-tensor}). Due to the anisotropic and nonsymmetric nature of the material, mirror diagonal symmetries ($\sigma_{d1}$ and $\sigma_{d2}$) are broken. The point group of the structure is downgraded to $C_4$, which is the group that has the $90^{\circ}$ rotation about the z-axis as its generator. The corresponding degenerate resonances split as their two-dimensional IRREP ($\Gamma_5$ of $C_{4v}$) splits into two one-dimensional IRREPs ($\Gamma_3$ and $\Gamma_4$ of $C_{4}$)~\cite{BICs-grupos} (see Supplementary Material~\cite{Supp}, Section 2). 
 
In addition, the introduction of losses leaves the structure's symmetry unperturbed but limits the quality factors of the quasi-BICs, as pointed out in~\cite{BIC-losses-destruction}. Moreover, for the same values of $\Delta$, the quality factor and amplitude of the resonance will drop significantly. Nevertheless, losses in the material make possible such physical effects as the MCD and isolation. 

\section{Coupled Mode Theory Description}

\begin{figure}[tb]
    \centering 
    \includegraphics[width=1\columnwidth]{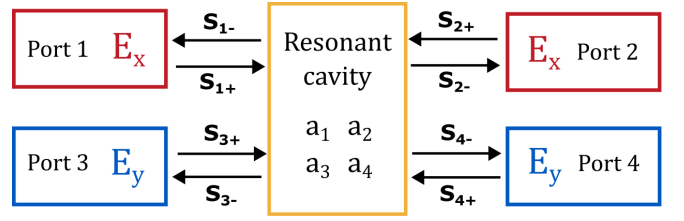}
    \caption{Schematics of the ports and the resonant metasurface for the CMT analysis. }
    \label{fig: Puertos}
\end{figure}

To better understand the physics behind the structure and as a guideline for the design of actual devices, we developed a theoretical model based on temporal CMT. More specifically, our model will consider the interaction of four ports (two ports for each linear polarization) and four resonant modes (two electric and two magnetic modes in the presence of the magnetic bias). Defining the ports of the structure as in Fig.~\ref{fig: Puertos}, where ports 1 and 2 accounts for the x-polarization channel and ports 3 and 4 form the y-polarization channel. The $C_4$ symmetry of the structure ensures that the scattering matrix (expressed in the linear-polarization basis) of the device has the form (Supplementary Material~\cite{Supp})

\begin{equation}\label{eq:S-matrix}
\mathbf{S}_{\rm lin}=
\begin{pmatrix}
 r & t& r_{xy} & t_{xy}\\
 t & r & t_{xy} & r_{xy}\\
 -r_{xy}& -t_{xy}& r & t\\
 -t_{xy}& -r_{xy}& t & r \\
\end{pmatrix}.
\end{equation}
where $r$ and $t$ represent the reflection and transmission amplitude coefficients when the polarization state of the incident light is preserved (co-polarized coefficients), while $r_{xy}$ and $t_{xy}$  are the reflection and transmission coefficients when the polarization is changed (cross-polarized coefficients). 
% In this case, the $C_4$ symmetry ensures that $t_{xy}=-t_{yx}$ and $r_{xy}=-r_{yx}$.
The scattering matrix of the metasurface in the circular-polarization basis with basis vectors $\mathbf{e}_{\pm}=(\mathbf{e}_x\mp j\mathbf{e}_y)/\sqrt{2}$ has the form (Supplementary Material~\cite{Supp})

\begin{equation}
  \mathbf{S}_{\rm circ}=
\begin{pmatrix}
    r_{\scalebox{0.75}[0.75]{L-R}}^{\rm f} & t_{\scalebox{0.75}[0.75]{L-L}}^{\rm b} & 0 & 0\\
    t_{\scalebox{0.75}[0.75]{R-R}}^{\rm f} & r_{\scalebox{0.75}[0.75]{R-L}}^{\rm b} & 0 & 0\\
    0 & 0 & r_{\scalebox{0.75}[0.75]{R-L}}^{\rm f} & t_{\scalebox{0.75}[0.75]{R-R}}^{\rm b}\\
    0 & 0 & t_{\scalebox{0.75}[0.75]{L-L}}^{\rm f} & r_{\scalebox{0.75}[0.75]{L-R}}^{\rm b}\\
\end{pmatrix}=\begin{pmatrix}
 r_{+}& t_{+}& 0& 0\\  
 t_{+}& r_{+}& 0& 0\\
 0& 0& r_{-}& t_{-}\\
 0& 0& t_{-}& r_{-}
\end{pmatrix},  \label{eq:S-matrix-xy}
\end{equation}
where the superscripts ``f" and ``b" indicate forward ($+z$) or backward propagation ($-z$), and the subscripts ``L'' and ``R'' indicate left and right-handedness of circular polarization, respectively. The elements of the scattering matrices in the two different polarization bases are related as: $t_{+}=t+jt_{xy}$, $r_{+}=r+jr_{xy}$, $t_{-}=t- jt_{xy}$ and $r_{-}=r-jr_{xy}$. %For completeness, in the left equation in Eq.~(\ref{eq:S-matrix-xy}), the elements have been properly identified~\cite{BICs-chirality}. 
It is worth mentioning that in the circular-polarization basis, nonreciprocity does not lead to a nonsymmetric scattering matrix due to the way that the elements of the matrix are defined, as discussed in~\cite{CMT-Chiral-Ports}. Moreover, as Eq.~\ref{eq:S-matrix-xy} shows, the notation chosen embraces the mathematical definition of the circular-polarization basis, uniquely defined by the sign. Light handedness implies a choice in nomenclature and point of view, as pointed out and widely discussed in~\cite{EM-chirality-Caloz}. Using the sign in the relative phase between components as the defining factor leaves no doubt about what we refer to. In any case, the convention used to relate both parts of Eq.~\ref{eq:S-matrix-xy} is the one commonly used in engineering, where the time dependence is defined as $e^{j\omega t}$ and the observer points towards the direction of propagation to declare the handedness.

The phenomenological CMT model is developed following the general coupled-mode equations~\cite{CMT-OG}:
\begin{gather}
    \frac{d\mathbf{a}}{dt}=(j\Omega-\Gamma)\mathbf{a}+K^T|s_+ \rangle \label{eq:CMT-1}, \\
    |s_- \rangle=C|s_+ \rangle+D\mathbf{a}\label{eq:CMT-2}.
\end{gather}
where  $\mathbf{a}$ is the  vector of $n \times 1$ size that represents the resonance amplitudes inside the metasurface and $n=4$ is the number of resonant modes. Here the resonance amplitudes are normalized such that $|a_i|^2$ 
corresponds to the energy stored in the $i$-th  resonant  mode. 
Moreover, in Eqs.~\ref{eq:CMT-1}--\ref{eq:CMT-2},
$D$ is the output  $m \times n$  coupling matrix,    $m=4$ is the number of ports,  $C$ is the scattering matrix of the background (or the direct pathway coupling) that is described as an $m\times m$ matrix, $K$ is the input coupling matrix of   $m\times n$ size, $\Gamma$ and $\Omega$ are $n\times n$ matrices which correspond to the decay rates and the resonance frequencies, respectively, and $|s_+ \rangle $ and $|s_- \rangle $ are vectors of $m \times 1$ size accounting for the output and input waves (see Fig.~\ref{fig: Puertos}).% \red{A $e^{\red{+j \omega t}}$ dependence in time is omitted. -- What is the meaning of this sentence?} 

There are certain conditions that the elements in Eqs.~\eqref{eq:CMT-1} and \eqref{eq:CMT-2} have to follow in order to preserve energy conservation~\cite{CMT-eqs}. The restrictions are as follows:

\begin{gather}
    C^{\dagger}C=I,\label{eq:Energy-conservation-1}\\
    D^{\dagger}D=K^{\dagger}K=2\Gamma,\label{eq:Energy-conservation-2}\\
    CK^{*}+D=0,\label{eq:Energy-conservation-3}\\
    C^TD^{*}+K=0.\label{eq:Energy-conservation-4}
\end{gather}

From the eigenmode analysis, we found that the four resonances are orthogonal and, therefore,  $\Gamma\Omega=\Omega\Gamma$. Both operators share eigenvalues and eigenvectors as they commute. Moreover, because of the in-plane distribution of electric and magnetic fields, the pair of magnetic resonances present an odd field symmetry with respect to the $z=0$ plane, while the pair of electric resonances have an even field symmetry. This will affect the coupling between the resonances and the excitation. Furthermore, it is known that the split in resonances is due to the off-diagonal components of the permittivity tensor, which now has circularly polarized waves as eigensolutions~\cite{Optical-isolator}. This will make the magnetic/electric resonance split into two modes: one that couples with RCP light and one with LCP light. To model this coupling, a relative $-\pi/2$ phase is introduced between $x-y$ components in the coupling matrices ($D$ and $K$) in resonances 2 and 4. And a relative $+\pi/2$ phase is introduced between $x-y$ components in resonances 1 and 3 (See Supplementary~\cite{Supp}, Section 4). 

In addition, the metasurface has geometric mirror symmetry with respect to the $z=0$ plane. The coupling amplitudes to the resonances in the metasurface have to be the same for the forward and backward illuminations. Due to  Eqs.~\ref{eq:Energy-conservation-2}--\ref{eq:Energy-conservation-4}, the elements of each of $K$ and $D$ matrices are related to one another. In terms of the $D$ matrix, its complex elements defined as $d_{ik}=d_{ik}e^{j\theta_{ik}}$ have the following relations: %\red{Please use begin array end array command for the multi-line equations. They work much better}
\begin{equation}\label{eq:module-1}
\begin{split}
    |d_{2k}|=|d_{1k}|,\\
    |d_{3k}|=|d_{4k}|. 
\end{split}    
\end{equation}

Another constraint is found for the phases~\cite{CMT-OG}, where the mirror symmetry applies a constraint of well-defined parity, as in our case, and the phases have to comply 
\begin{equation}\label{eq:angle-1}
\begin{split}
    \theta_{1j}=\theta_{2j}+2\pi N \quad \text{or}\quad \theta_{1j}=\theta_{2j}+(2N+1)\pi,\\
    \theta_{3j}=\theta_{4j}+2\pi N \quad \text{or}\quad \theta_{3j}=\theta_{4j}+(2N+1)\pi,    
\end{split}
\end{equation}
where $N$ is an integer. In the considered case, resonances 1 and 2 are even (the magnetic dipole resonances), while resonances 3 and 4 are odd (the electric dipole resonances).

Matrices $\Omega$ and $\Gamma$ are assumed to be diagonal, with components $\Omega= {\rm diag} (\omega_m^-,\omega_m^+,\omega_e^-,\omega_e^+)$ and $\Gamma= {\rm diag} (\Gamma_m^-,\Gamma_m^+,\Gamma_e^-,\Gamma_e^+)$, where the subscript refers to the nature of the resonance and its symmetry (magnetic and electric, respectively) and the superscript relates to which circular polarization the resonance couples after splitting due to the external magnetic field. In addition, a canonical phenomenological way to introduce losses in the model is to decompose the $\Gamma$ matrix as $\Gamma=\gamma+1/\tau$, where the first matrix refers to the decay rates of the resonances coupling into the ports and the second matrix reflects the material losses. A simple and standard way to solve Eqs. \eqref{eq:CMT-1} and \eqref{eq:CMT-2} is first to solve the system of equations neglecting the material losses with all the previous constraints ($1/\tau =0$) and then introducing the change that accounts for the material loss as $\Gamma_{m,e}^{\pm}=\gamma_{m,e}^{\pm}+1/\tau_{m,e}^{\pm}$. As a result, we obtain the final expressions for the $S$-matrix coefficients in the circular-polarization basis: 
\begin{gather}\label{eq:CMT-solutions}
\begin{split}
    t_+ = t_d
        -\frac{\gamma_m^+(r_d+ t_d)}{j(\omega-\omega_m^+)+\gamma_{m}^++1/\tau_{m}^+}\\
        +\frac{\gamma_e^+(r_d- t_d)}{j(\omega-\omega_e^+)+\gamma_{e}^++1/\tau_{e}^+},
\end{split} \\
\begin{split}
    t_- = t_d 
        -\frac{\gamma_m^-(r_d+ t_d)}{j(\omega-\omega_m^-)+\gamma_{m}^-+1/\tau_{m}^-} \\
        +\frac{\gamma_e^-(r_d- t_d)}{j(\omega-\omega_e^-)+\gamma_{e}^-+1/\tau_{e}^-},
\end{split} \\
\begin{split}
    r_+ = r_d
        -\frac{\gamma_m^+(r_d+ t_d)}{j(\omega-\omega_m^+)+\gamma_{m}^++1/\tau_{m}^+}\\
        -\frac{\gamma_e^+(r_d- t_d)}{j(\omega-\omega_e^+)+\gamma_{e}^++1/\tau_{e}^+},
\label{eq:CMT-solutions33}
\end{split} \\
\begin{split}
    r_- = r_d
        -\frac{\gamma_m^-(r_d+ t_d)}{j(\omega-\omega_m^-)+\gamma_{m}^-+1/\tau_{m}^-}\\
        -\frac{\gamma_e^-(r_d- t_d)}{j(\omega-\omega_e^-)+\gamma_{e}^-+1/\tau_{e}^-}.\label{eq:CMT-solutions4}
\end{split} 
\end{gather}
In the derivation, the direct pathway (background) scattering matrix was modeled as
\begin{equation}C=
    \begin{pmatrix}
    r_{d} & t_{d} & 0 & 0\\
    t_{d} & r_{d} & 0 & 0\\
    0 & 0 & r_{d} & t_{d}\\
    0 & 0 & t_{d} & r_{d}
    \end{pmatrix},
\end{equation}

where $t_d$ and $r_d$ are the transmission and reflection coefficients, respectively. The background pathway has been considered reciprocal and without losses, i.e., the $C$ matrix is a unitary matrix.

%-----------------------------------------------------------------------------------

\section{Maximization of Nonreciprocity and Magnetic Circular Dichroism}
\begin{figure*}[t!]
    \centering
    \includegraphics[width=1\textwidth]{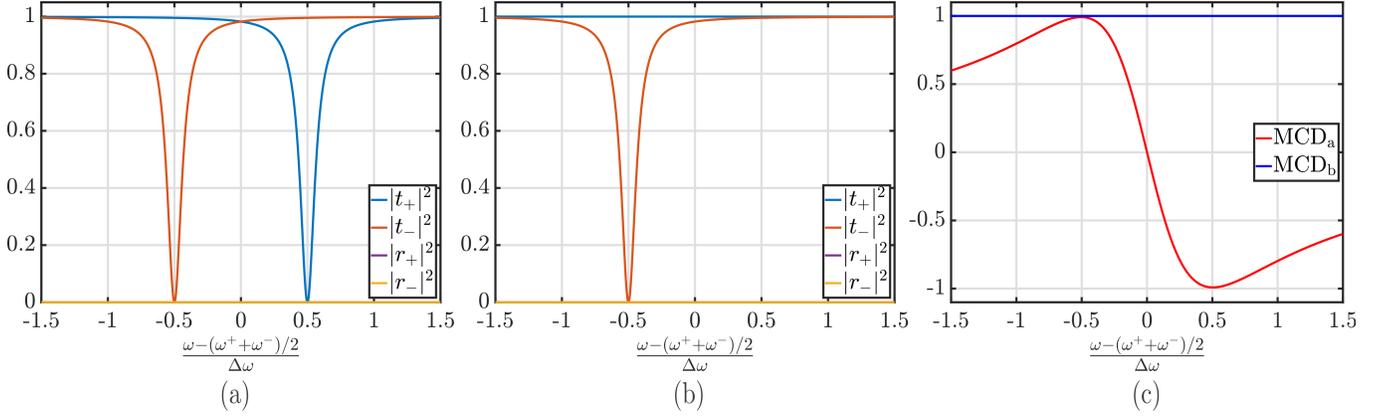}
    \caption{(a) Reflectances and transmittances for an ideal metasurface exhibiting maximal isolation for circularly polarized light of a given handedness. The metasurface behaves as a broadband-matched resonant absorber where the resonant frequency depends on the illumination direction. 
    In this example, $\gamma^+ = \gamma^-=1/\tau^+ = 1/\tau^-= \gamma $, $t_{\rm d}=1$, and  $\Delta\omega=\omega^+-\omega^-=30\gamma$. The horizontal axis is normalized in such a way that its center corresponds to the average between $\omega^+$ and $\omega^-$. At frequency $\omega^+$, transmission for the backward propagating RCP light reaches $|t_-(\omega_+)|^2=0.996$.
    % (b)   The values for the used variables are $\omega_e^+=\omega_m^+=\omega^+$, $\omega_e^-=\omega_m^-=\omega^-$, $\gamma_m^+=\gamma_m^-=\gamma_e^+=\gamma_e^-=\gamma$ and $1/\tau_m^+=1/\tau_m^-=1/\tau_e^+=1/\tau_e^-=\gamma$, with $\Delta\omega=\omega^+-\omega^-=30\gamma$, $t_d=1$. 
    (b) The same for a metasurface that exhibits no material loss for RCP (LCP) light when illuminated in the backward (forward) direction. (c) MCD for the two limiting scenarios in (a) and (b).   }
    \label{fig:th}
\end{figure*}

In this section, we derive the ideal conditions under which the metasurface with the aforementioned properties provides the highest (unitary) Magnetic Circular Dichroism (MCD) and the highest isolation. MCD is conventionally defined in terms of the difference in absorptions for the two circular polarizations of light~\cite{MCD-def}:
\begin{equation}
    \rm MCD=\frac{A_{-}-A_{+}}{A_{-}+A_{+}}= \frac{(|t_+|^2-|t_-|^2)+(|r_+|^2-|r_-|^2)}{2-(|t_-|^2+|t_+|^2)-(|r_-|^2+|r_+|^2)}.\label{eq:mcdl}
\end{equation}
where $A_{\pm}=1-|t_{\pm}|^2-|r_{\pm}|^2$. In addition, we define the isolation ratio as $\iota = |t_-/t_+|$.
In the case at hand, achieving a high isolation ratio will lead to high MCD, as seen from Eq.~\ref{eq:S-matrix-xy}. For instance,  when illuminating one side (the "$+z$" direction) with RCP light, the response is $t_+$, while LCP light yields $t_-$. Conversely, illuminating from the opposite side (``$-z$'' direction) with RCP light produces $t_-$, and with LCP, $t_+$. A high MCD means one polarization is mainly absorbed, and the other is not; the previously blocked polarization is not absorbed when the direction is reversed. 
Next, we look for the conditions to design a complete isolator for circularly polarized light, that is, a device that fully absorbs one polarization from one side and transmits it from the other.

We start by finding the conditions to achieve high isolation  for circularly polarized light  using the theoretical results obtained in the last section, that is, Eqs.~\ref{eq:CMT-solutions}-\ref{eq:CMT-solutions4}. Thus, we aim to ensure the full absorption and full transmission of light of a given circular polarization incident on the metasurface from the two opposite directions. Without loss of generality, we determine these two conditions for the right-handed circular polarization (as shown below, the dual conditions will be automatically satisfied for the opposite polarization). Mathematically, the conditions read as 
\begin{equation}\label{conditionsvik}
   A^{\rm f}_{\rm R}=1-|t_{\rm R-R}^{\rm f}|^2 -|r_{\rm L-R}^{\rm f}|^2=1, \quad |t_{\rm R-R}^{\rm b}| =1.
\end{equation}
Naturally, they would lead to the scenario of   maximization of the   $\iota$ ratio.
To simplify the derivation, in the following, the case where $r_d=0$ and $t_d=1$ is explored. 
% We want to point out that the equations for the plus and minus polarizations have the same mathematical form. Therefore, we develop the conditions for only one polarization in what follows. The results for the other polarization can be trivially obtained by interchanging superscripts and subscripts $(+)\leftrightarrow(-)$. 
Using Eqs.~\ref{eq:CMT-solutions}-\ref{eq:CMT-solutions4} and Eq.~\ref{eq:S-matrix-xy}, we can express absorption of the RCP light incident in the forward direction     as $A^{\rm f}_{\rm R}= 1-|t_{+}|^2-|r_{+}|^2= A_+ $, which results in 
\begin{equation}
\begin{split}\label{eq:Absorption}
    \rm{ A_+} = \frac{2\gamma_m^+/\tau_m^+}{(\gamma_m^++1/\tau_m^+)^2+(\omega-\omega_m^+)^2}\\
    +\frac{2\gamma_e^+/\tau_e^+}{(\gamma_e^++1/\tau_e^+)^2+(\omega-\omega_e^+)^2}.  
\end{split}
\end{equation}
To satisfy the first condition (maximization of $\rm{ A_+} $), the eigenfrequencies for the resonances of both parities have to be equal $\omega_m^+=\omega_e^+=\omega^+$, and the maximum will take place at the frequency $\omega=\omega^+$. To reach precisely unitary absorption, we should ensure $|r_{+}|^2=0$ and $|t_{+}|^2=0$. We study both cases separately and then combine the conditions. At the resonant frequency, $|r_{+}|^2=0$ is obtained if
\begin{equation}\label{eq:r0}
    \gamma_{m}^{+}\tau_{m}^{+}= \gamma_{e}^{+}\tau_{e}^{+}.
\end{equation}
This constraint   leads to  the condition of a Huygens' metasurface~\cite{Huygens-review}. Furthermore, forcing separately that $|t_{+}|^2=0$ at the resonance frequency will yield 
\begin{equation}\label{eq:t0}
    \frac{1}{\tau_{m}^{+}}\frac{1}{\tau_{e}^{+}}= \gamma_{m}^{+}\gamma_{e}^{+}.
\end{equation}
Combining both Eqs.~\ref{eq:r0}-\ref{eq:t0}, we end up with the know condition of critical coupling that ensures full absorption of incoming energy at the resonance frequency:
\begin{equation}\label{eq:Critical-coupling}
   \frac{1}{\tau_{m}^{+}} =\gamma_{m}^{+}, \qquad \frac{1}{\tau_{e}^{+}} =\gamma_{e}^{+}. 
\end{equation}
 This result is widely known for reciprocal metasurfaces and can be interpreted as a balance between the losses induced by the material and the scattering losses~\cite{CMT-Absorption-condition-1}. In contrast, in our case, the critical coupling condition   holds at a given frequency only for one illumination direction. 
 Substituting conditions~\ref{eq:Critical-coupling} into Eq.~\ref{eq:Absorption}, we indeed obtain $A_+=1$.

In what follows, we additionally require  zero reflection condition $|r_{+} (\omega)|^2=|r_{-} (\omega)|^2=0$ for all frequencies $\omega$. Such a broadband Huygens' regime~\cite{Broadband-Huygens} has multiple practical advantages such as the possibility to cascade several metasurfaces and the minimization of parasitic interference.
As is seen from Eq.~\ref{eq:CMT-solutions33}, broadband suppression of reflections for illuminations from both sides  is achieved when we choose $\omega_m^-=\omega_e^-= \omega^-$ and  additionally satisfy: 
\begin{equation}\label{eq:viktaradded}
\begin{split}
\gamma_{m}^{+}= \gamma_{e}^{+}= \gamma^{+}, \quad \tau_{m}^{+}=  \tau_{e}^{+}=\tau^{+},  \\
\gamma_{m}^{-}= \gamma_{e}^{-}=\gamma^{-}, \quad \tau_{m}^{-}=  \tau_{e}^{-}=\tau^{-}.    
\end{split}
\end{equation}

Next, to satisfy the second condition in Eq.~\ref{conditionsvik} at a given frequency $\omega$, there are two possible \textit{limiting} scenarios. The first one is to ensure full absorption for  RCP light incident on the metasurface in the backward direction $A^{\rm b}_{\rm R}= 1-|t_{\rm R-R}^{\rm b}|^2 -|r_{\rm L-R}^{\rm b}|^2 = A_-$ but occurring at a different frequency $\omega^- \neq \omega^+$. By decoupling the two resonance frequencies such that the spectral distance between them is much greater than their widths, i.e., $|\omega^- - \omega^+| \gg |\gamma_{e,m}^+ +1/\tau_{e,m}^+|$, we can asymptotically reach $|t_{\rm R-R}^{\rm b}| \approx 1$ at the frequency of $\omega^+$. This scenario is depicted in Fig.~\ref{fig:th}(a). Due to the condition in Eq.~\ref{eq:viktaradded}, reflections are zero at all frequencies, while the absorption peaks for two light propagation directions are spectrally separated. To have complete absorption   $A_-(\omega^-)=1$, one needs to additionally satisfy 
$1/\tau^{-} =\gamma^{-}$.  

The second limiting scenario for satisfying the condition in~\ref{conditionsvik} is when the metasurface is completely transparent at all frequencies  (unitary transmission with an arbitrary phase) for RCP light incident in the backward direction. This scenario  is presented in Fig.~\ref{fig:th}(b).  It can be achieved if $1/\tau^-=0$.

As was mentioned above, at the frequencies where  the  isolation ratio $\iota=|t_-/t_+|$ is high, the MCD is large.  
In Fig.~\ref{fig:th}(c), we plot the MCD parameter for the two above scenarios. While for the first scenario, the MCD approaches to unit value at $\omega = \omega^+$, in the second scenario, it is a unit for all the frequencies due to its definition.

\section{Ferrite metasurface realization}

\begin{figure*}[t]
    \centering
    \includegraphics[width = 1\textwidth]{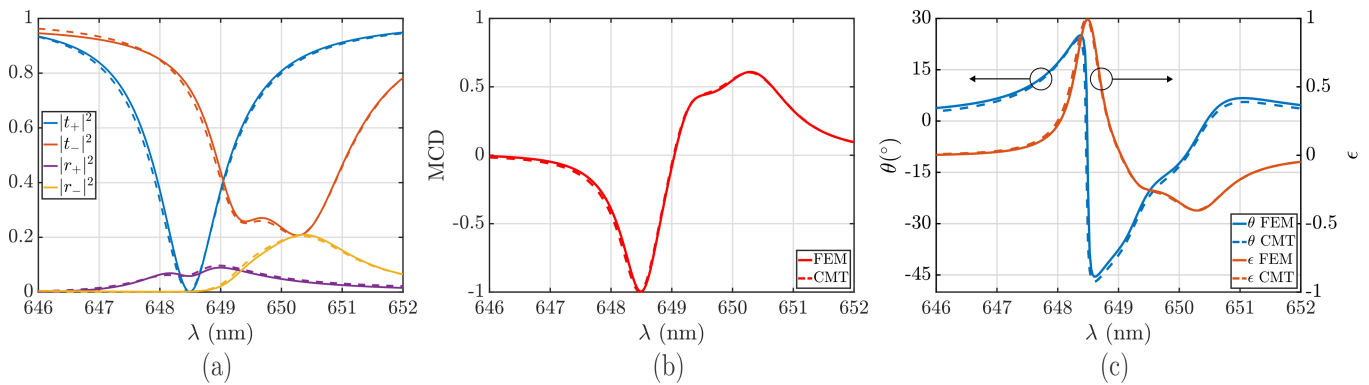}
    \caption{(a) Reflectances and transmittances of the optimized metasurface when illuminated with circularly polarized light at normal incidence. The CMT fit is plotted with a dashed line with the same color as the simulated data (denoted as FEM). 
    (b) MCD of the metasurface. (c)  The Faraday rotation angle and ellipticity of the designed metasurface versus the wavelength.
    The geometrical parameters of the structure are $w=430$ nm, $h=135.6$ nm, $l=269.2$ nm, and $\Delta=24.7$ nm. The values obtained for the CMT fit are $\omega_m^-=2896.3$ THz, $\omega_m^+=2904.8$ THz, $\omega_e^-=2901.5$ THz and $\omega_e^+=2904.8$ THz for the eigenfrequencies; $\gamma_m^-=\gamma_m^+=\gamma_{m}=2.2$ THz, $\gamma_e^-=\gamma_e^+=\gamma_{e}=0.8$ THz for the scattering decay rates; $1/\tau_m^-=2.25$ THz, $1/\tau_e^-=1.69$ THz and $1/\tau_m^+=1/\tau_e^+=1.42$ THz for the absorption decay rates. Finally, $t_d=0.998$ is used as the direct pathway parameter.}
    \label{fig:CD_optim}
\end{figure*}
%\red{Your (b) and (c) must be interchanged. see the order in the caption. Luis, please combine current subfigures (b) and (c) into one subfigure and use left and right different vertical axes for that. Add actual MCD to fig. 6 as a new subfigure (c).}

Using a metasurface with the geometry described in Fig.~\ref{fig: Geometry} and ferrite material (${\rm Bi3YIG}$~\cite{Meta-BiYIG-material}) with a tensor given by Eq.~\ref{eq:permittivity-tensor}, the maximization of MCD and isolation properties studied theoretically in the last section can be applied. In the case of the real structure, the mentioned configuration for the resonances must be achieved by tweaking the geometrical parameters. Studying the behavior of the structure when there is no bias and no losses ($\varepsilon_{xy}=0$ and $\varepsilon_{xx}=\varepsilon_{1r}$), it is apparent that the decay rates of the magnetic and electric resonances are intrinsically different, see Fig.~\ref{fig:BICs_Delta_Nobias_noloss}. However, both can be changed by tuning the asymmetry parameter $\Delta$. It must also be noted that the wavelength shift of the electric dipole resonances is noticeable when $\Delta$ is changed. In contrast, the magnetic dipole resonances shift much slower,  as is seen in Fig.~\ref{fig:BICs_Delta_Nobias_noloss}(c) and Fig.~\ref{fig:Linear-biased}. 
The bias will split each of the magnetic and electric dipole resonances into two, one coupling to the plus sign of the helical basis vectors and one for the minus sign. By introducing losses, the behavior of the resonances holds, but the resonances suffer a fall in quality factor and do shift. The final design uses optimization techniques for the geometrical parameters and the aforementioned knowledge to maximize the MCD. It should be noted that due to the material restrictions, we could not reach either of the two limiting scenarios shown in Figs.~\ref{fig:th}(a) and (b). Nevertheless,   characteristics of the designed metasurface are closer to those in Figs.~\ref{fig:th}(a).

The simulated scattering parameters of the metasurface, when illuminated normally with circularly polarized light, are shown in Fig.~\ref{fig:CD_optim}(a). The CMT fitting is also shown in the same plot, accurately modeling the simulated data.  A maximum for the MCD is obtained at $\lambda=648.5$ nm, reaching $\rm{|MCD|}=0.996$, as shown in Fig.~\ref{fig:CD_optim}(b). At the same wavelength, $|t_+|$ is almost fully suppressed. Moreover, the amount of energy transmitted forward with LCP polarization reaches $73.90\%$ while only a $0.12\%$ of LCP light would be transmitted in the backward direction. The raw contrast between transmittances $|t_{\scalebox{0.75}[0.75]{L-L}}^{\rm f}/t_{\scalebox{0.75}[0.75]{L-L}}^{\rm b}|^2= |t_-/t_+|^2= \iota^2 \approx  625$, which can be expressed as a 27.89~dB isolation ratio.
The energy with a defined circular polarization can go through the device in one direction while being absorbed in the opposite direction. 

Finally,  in Fig.~\ref{fig:CD_optim}(c), we plot the Faraday rotation angle $\theta$ and ellipticity $\epsilon$ versus frequency for the designed metasurface. We use the conventional definition for the rotation angle $\theta = 1/2\arctan{[2\Re{(\chi)}/(1-|\chi|^2)]}$, where $\chi=E^{\rm tr}_y/E^{\rm tr}_x$. Where $\Re$ denotes the real part of a function. The ellipticity $\epsilon$ is defined as the ratio between the size of the major and the minor axis of the polarization ellipse of transmitted light. At $\lambda=648.5$ nm the ellipticity reaches almost unity. This occurs because almost all of the transmitted light is circularly polarized with left-handedness. % \red{You define Faraday rotation as $\theta$ but you do not define any character for the ellipticity. Looks strange. also in the plots. let us use some variable. e.g., $\epsilon$}
On the other hand, at $\lambda=649.05$ nm where ellipticity is equal to zero,  the metasurface rotates linearly polarized light at an angle $\theta=36.4^{\circ}$ with the  transmittance $|t|^2 = 40.5\% $.
The obtained values of the rotation angle and ellipticity are giant, taking into account the subwavelength thickness of the metasurface ($h/\lambda=0.21$) and the realistic material parameters used. We can also estimate the effective Verdet constant of the metasurface $V = \theta /(\mu_0 H_0 h)$~\cite{Nonreciprocity}. Taking $\mu_0 H_0=0.66$~T from \cite{Meta-BiYIG-material}, the Verdet constant reaches the maximum value of $V=7.12 \times 10^6$~rad/(${\rm T \cdot m}$), which exceeds the value for thin-film of conventional MO materials by 3--4 orders of magnitude~\cite{Nonreciprocity}. 
Such giant enhancement of the MO properties is due to  the use of quasi-BICs in our geometry.

\section{Conlcusions}

We have developed a metasurface made out of a magnetic ferrite and based on quasi-BIC resonances that, with the use of orthogonal pairs of resonances with electric and magnetic nature, can achieve high MCD combined with strong  isolation for circular polarization. The device acts with this configuration as a one-way isolator, as the transmission for the non-blocked polarization is maximized. In addition, by matching the resonance amplitudes and phases, it is possible to use the quasi-BIC-empowered metasurface to enhance the Faraday effect of the structure. An enormous figure of merit is obtained using this approach. All of the above has been designed using a realistic material (${\rm Bi3YIG}$), experimentally characterized, and with noticeable losses. 

The finite-element-method simulations have been supported with a coupled mode theory description of the metasurface. This phenomenological theoretical model has been useful in predicting and explaining the capabilities of the device. Furthermore, it supports the observed enhancement of the Faraday effect and proves the use of BICs as a path to control and enhance MO effects. The realization of a Faraday rotator can lead to the creation of an ultra-compact isolator using two linear polarizers in the famous Faraday isolator architecture. 

In order to design a polarization-insensitive optical isolator, one needs to additionally break the inversion (parity) symmetry that is currently present in the metasurface design. This can be straightforwardly achieved using a conventional approach by introducing two circular polarizers (with the same handedness) around the metasurface (e.g., the Supplementary Information in Ref.~\cite{Geometry-1} and Section 6 in Supplementary material~\cite{Supp}).

\section*{Acknowledgment}
This research was supported by the Beatriz Galindo excellence grant (grant No. BG-00024), the Spanish National Research Council (grant No. PID2021-128442NA-I00), and   the Academy of Finland (Project No. 356797).

\newpage

\bibliography{bibliography}

%apsrev4-2.bst 2019-01-14 (MD) hand-edited version of apsrev4-1.bst
%Control: key (0)
%Control: author (8) initials jnrlst
%Control: editor formatted (1) identically to author
%Control: production of article title (0) allowed
%Control: page (0) single
%Control: year (1) truncated
%Control: production of eprint (0) enabled
\begin{thebibliography}{52}%
\makeatletter
\providecommand \@ifxundefined [1]{%
 \@ifx{#1\undefined}
}%
\providecommand \@ifnum [1]{%
 \ifnum #1\expandafter \@firstoftwo
 \else \expandafter \@secondoftwo
 \fi
}%
\providecommand \@ifx [1]{%
 \ifx #1\expandafter \@firstoftwo
 \else \expandafter \@secondoftwo
 \fi
}%
\providecommand \natexlab [1]{#1}%
\providecommand \enquote  [1]{``#1''}%
\providecommand \bibnamefont  [1]{#1}%
\providecommand \bibfnamefont [1]{#1}%
\providecommand \citenamefont [1]{#1}%
\providecommand \href@noop [0]{\@secondoftwo}%
\providecommand \href [0]{\begingroup \@sanitize@url \@href}%
\providecommand \@href[1]{\@@startlink{#1}\@@href}%
\providecommand \@@href[1]{\endgroup#1\@@endlink}%
\providecommand \@sanitize@url [0]{\catcode `\\12\catcode `\$12\catcode
  `\&12\catcode `\#12\catcode `\^12\catcode `\_12\catcode `\%12\relax}%
\providecommand \@@startlink[1]{}%
\providecommand \@@endlink[0]{}%
\providecommand \url  [0]{\begingroup\@sanitize@url \@url }%
\providecommand \@url [1]{\endgroup\@href {#1}{\urlprefix }}%
\providecommand \urlprefix  [0]{URL }%
\providecommand \Eprint [0]{\href }%
\providecommand \doibase [0]{https://doi.org/}%
\providecommand \selectlanguage [0]{\@gobble}%
\providecommand \bibinfo  [0]{\@secondoftwo}%
\providecommand \bibfield  [0]{\@secondoftwo}%
\providecommand \translation [1]{[#1]}%
\providecommand \BibitemOpen [0]{}%
\providecommand \bibitemStop [0]{}%
\providecommand \bibitemNoStop [0]{.\EOS\space}%
\providecommand \EOS [0]{\spacefactor3000\relax}%
\providecommand \BibitemShut  [1]{\csname bibitem#1\endcsname}%
\let\auto@bib@innerbib\@empty
%</preamble>
\bibitem [{\citenamefont {Quevedo-Teruel}\ \emph {et~al.}(2019)\citenamefont
  {Quevedo-Teruel}, \citenamefont {Chen}, \citenamefont {D{\'i}az-Rubio},
  \citenamefont {Gok}, \citenamefont {Grbic}, \citenamefont {Minatti},
  \citenamefont {Martini}, \citenamefont {Maci}, \citenamefont {Eleftheriades},
  \citenamefont {Chen}, \citenamefont {Zheludev}, \citenamefont {Papasimakis},
  \citenamefont {Choudhury}, \citenamefont {Kudyshev}, \citenamefont {Saha},
  \citenamefont {Reddy}, \citenamefont {Boltasseva}, \citenamefont {Shalaev},
  \citenamefont {Kildishev}, \citenamefont {Sievenpiper}, \citenamefont
  {Caloz}, \citenamefont {Al}, \citenamefont {He}, \citenamefont {Zhou},
  \citenamefont {Valerio}, \citenamefont {Rajo-Iglesias}, \citenamefont
  {Sipus}, \citenamefont {Mesa}, \citenamefont {Rodr{\'i}guez-Berral},
  \citenamefont {Medina}, \citenamefont {Asadchy}, \citenamefont {Tretyakov},\
  and\ \citenamefont {Craeye}}]{Roadmap-meta}%
  \BibitemOpen
  \bibfield  {author} {\bibinfo {author} {\bibfnamefont {O.}~\bibnamefont
  {Quevedo-Teruel}}, \bibinfo {author} {\bibfnamefont {H.}~\bibnamefont
  {Chen}}, \bibinfo {author} {\bibfnamefont {A.}~\bibnamefont
  {D{\'i}az-Rubio}}, \bibinfo {author} {\bibfnamefont {G.}~\bibnamefont {Gok}},
  \bibinfo {author} {\bibfnamefont {A.}~\bibnamefont {Grbic}}, \bibinfo
  {author} {\bibfnamefont {G.}~\bibnamefont {Minatti}}, \bibinfo {author}
  {\bibfnamefont {E.}~\bibnamefont {Martini}}, \bibinfo {author} {\bibfnamefont
  {S.}~\bibnamefont {Maci}}, \bibinfo {author} {\bibfnamefont {G.~V.}\
  \bibnamefont {Eleftheriades}}, \bibinfo {author} {\bibfnamefont
  {M.}~\bibnamefont {Chen}}, \bibinfo {author} {\bibfnamefont {N.~I.}\
  \bibnamefont {Zheludev}}, \bibinfo {author} {\bibfnamefont {N.}~\bibnamefont
  {Papasimakis}}, \bibinfo {author} {\bibfnamefont {S.}~\bibnamefont
  {Choudhury}}, \bibinfo {author} {\bibfnamefont {Z.~A.}\ \bibnamefont
  {Kudyshev}}, \bibinfo {author} {\bibfnamefont {S.}~\bibnamefont {Saha}},
  \bibinfo {author} {\bibfnamefont {H.}~\bibnamefont {Reddy}}, \bibinfo
  {author} {\bibfnamefont {A.}~\bibnamefont {Boltasseva}}, \bibinfo {author}
  {\bibfnamefont {V.~M.}\ \bibnamefont {Shalaev}}, \bibinfo {author}
  {\bibfnamefont {A.~V.}\ \bibnamefont {Kildishev}}, \bibinfo {author}
  {\bibfnamefont {D.}~\bibnamefont {Sievenpiper}}, \bibinfo {author}
  {\bibfnamefont {C.}~\bibnamefont {Caloz}}, \bibinfo {author} {\bibfnamefont
  {A.}~\bibnamefont {Al}}, \bibinfo {author} {\bibfnamefont {Q.}~\bibnamefont
  {He}}, \bibinfo {author} {\bibfnamefont {L.}~\bibnamefont {Zhou}}, \bibinfo
  {author} {\bibfnamefont {G.}~\bibnamefont {Valerio}}, \bibinfo {author}
  {\bibfnamefont {E.}~\bibnamefont {Rajo-Iglesias}}, \bibinfo {author}
  {\bibfnamefont {Z.}~\bibnamefont {Sipus}}, \bibinfo {author} {\bibfnamefont
  {F.}~\bibnamefont {Mesa}}, \bibinfo {author} {\bibfnamefont {R.}~\bibnamefont
  {Rodr{\'i}guez-Berral}}, \bibinfo {author} {\bibfnamefont {F.}~\bibnamefont
  {Medina}}, \bibinfo {author} {\bibfnamefont {V.}~\bibnamefont {Asadchy}},
  \bibinfo {author} {\bibfnamefont {S.}~\bibnamefont {Tretyakov}},\ and\
  \bibinfo {author} {\bibfnamefont {C.}~\bibnamefont {Craeye}},\ }\bibfield
  {title} {\bibinfo {title} {Roadmap on metasurfaces},\ }\bibfield  {journal}
  {\bibinfo  {journal} {Journal of Optics (United Kingdom)}\ }\textbf {\bibinfo
  {volume} {21}},\ \href {https://doi.org/10.1088/2040-8986/ab161d}
  {10.1088/2040-8986/ab161d} (\bibinfo {year} {2019})\BibitemShut {NoStop}%
\bibitem [{\citenamefont {Singh}\ \emph {et~al.}(2022)\citenamefont {Singh},
  \citenamefont {Ahmed},\ and\ \citenamefont {Esselle}}]{Review-1}%
  \BibitemOpen
  \bibfield  {author} {\bibinfo {author} {\bibfnamefont {K.}~\bibnamefont
  {Singh}}, \bibinfo {author} {\bibfnamefont {F.}~\bibnamefont {Ahmed}},\ and\
  \bibinfo {author} {\bibfnamefont {K.}~\bibnamefont {Esselle}},\ }\bibfield
  {title} {\bibinfo {title} {Electromagnetic metasurfaces: Insight into
  evolution, design and applications},\ }\bibfield  {journal} {\bibinfo
  {journal} {Crystals}\ }\textbf {\bibinfo {volume} {12}},\ \href
  {https://doi.org/10.3390/cryst12121769} {10.3390/cryst12121769} (\bibinfo
  {year} {2022})\BibitemShut {NoStop}%
\bibitem [{\citenamefont {Ding}\ \emph {et~al.}(2018)\citenamefont {Ding},
  \citenamefont {Pors},\ and\ \citenamefont {Bozhevolnyi}}]{Review-2}%
  \BibitemOpen
  \bibfield  {author} {\bibinfo {author} {\bibfnamefont {F.}~\bibnamefont
  {Ding}}, \bibinfo {author} {\bibfnamefont {A.}~\bibnamefont {Pors}},\ and\
  \bibinfo {author} {\bibfnamefont {S.~I.}\ \bibnamefont {Bozhevolnyi}},\
  }\bibfield  {title} {\bibinfo {title} {Gradient metasurfaces: A review of
  fundamentals and applications},\ }\bibfield  {journal} {\bibinfo  {journal}
  {Reports on Progress in Physics}\ }\textbf {\bibinfo {volume} {81}},\ \href
  {https://doi.org/10.1088/1361-6633/aa8732} {10.1088/1361-6633/aa8732}
  (\bibinfo {year} {2018})\BibitemShut {NoStop}%
\bibitem [{\citenamefont {Chen}\ \emph {et~al.}(2016)\citenamefont {Chen},
  \citenamefont {Taylor},\ and\ \citenamefont {Yu}}]{Review-3}%
  \BibitemOpen
  \bibfield  {author} {\bibinfo {author} {\bibfnamefont {H.~T.}\ \bibnamefont
  {Chen}}, \bibinfo {author} {\bibfnamefont {A.~J.}\ \bibnamefont {Taylor}},\
  and\ \bibinfo {author} {\bibfnamefont {N.}~\bibnamefont {Yu}},\ }\bibfield
  {title} {\bibinfo {title} {A review of metasurfaces: Physics and
  applications},\ }\bibfield  {journal} {\bibinfo  {journal} {Reports on
  Progress in Physics}\ }\textbf {\bibinfo {volume} {79}},\ \href
  {https://doi.org/10.1088/0034-4885/79/7/076401}
  {10.1088/0034-4885/79/7/076401} (\bibinfo {year} {2016})\BibitemShut
  {NoStop}%
\bibitem [{\citenamefont {Jalas}\ \emph {et~al.}(2013)\citenamefont {Jalas},
  \citenamefont {Petrov}, \citenamefont {Eich}, \citenamefont {Freude},
  \citenamefont {Fan}, \citenamefont {Yu}, \citenamefont {Baets}, \citenamefont
  {Popovi{\'c}}, \citenamefont {Melloni}, \citenamefont {Joannopoulos},
  \citenamefont {Vanwolleghem}, \citenamefont {Doerr},\ and\ \citenamefont
  {Renner}}]{Isolator}%
  \BibitemOpen
  \bibfield  {author} {\bibinfo {author} {\bibfnamefont {D.}~\bibnamefont
  {Jalas}}, \bibinfo {author} {\bibfnamefont {A.}~\bibnamefont {Petrov}},
  \bibinfo {author} {\bibfnamefont {M.}~\bibnamefont {Eich}}, \bibinfo {author}
  {\bibfnamefont {W.}~\bibnamefont {Freude}}, \bibinfo {author} {\bibfnamefont
  {S.}~\bibnamefont {Fan}}, \bibinfo {author} {\bibfnamefont {Z.}~\bibnamefont
  {Yu}}, \bibinfo {author} {\bibfnamefont {R.}~\bibnamefont {Baets}}, \bibinfo
  {author} {\bibfnamefont {M.}~\bibnamefont {Popovi{\'c}}}, \bibinfo {author}
  {\bibfnamefont {A.}~\bibnamefont {Melloni}}, \bibinfo {author} {\bibfnamefont
  {J.~D.}\ \bibnamefont {Joannopoulos}}, \bibinfo {author} {\bibfnamefont
  {M.}~\bibnamefont {Vanwolleghem}}, \bibinfo {author} {\bibfnamefont {C.~R.}\
  \bibnamefont {Doerr}},\ and\ \bibinfo {author} {\bibfnamefont
  {H.}~\bibnamefont {Renner}},\ }\bibfield  {title} {\bibinfo {title} {What
  is-and what is not-an optical isolator},\ }\href
  {https://doi.org/10.1038/nphoton.2013.185} {\bibfield  {journal} {\bibinfo
  {journal} {Nature Photonics}\ }\textbf {\bibinfo {volume} {7}},\ \bibinfo
  {pages} {579} (\bibinfo {year} {2013})}\BibitemShut {NoStop}%
\bibitem [{\citenamefont {Asadchy}\ \emph
  {et~al.}(2020{\natexlab{a}})\citenamefont {Asadchy}, \citenamefont
  {Mirmoosa}, \citenamefont {D{\'i}az-Rubio}, \citenamefont {Fan},\ and\
  \citenamefont {Tretyakov}}]{Nonreciprocity}%
  \BibitemOpen
  \bibfield  {author} {\bibinfo {author} {\bibfnamefont {V.~S.}\ \bibnamefont
  {Asadchy}}, \bibinfo {author} {\bibfnamefont {M.~S.}\ \bibnamefont
  {Mirmoosa}}, \bibinfo {author} {\bibfnamefont {A.}~\bibnamefont
  {D{\'i}az-Rubio}}, \bibinfo {author} {\bibfnamefont {S.}~\bibnamefont
  {Fan}},\ and\ \bibinfo {author} {\bibfnamefont {S.~A.}\ \bibnamefont
  {Tretyakov}},\ }\bibfield  {title} {\bibinfo {title} {Tutorial on
  electromagnetic nonreciprocity and its origins},\ }\href
  {https://doi.org/10.1109/JPROC.2020.3012381} {\bibfield  {journal} {\bibinfo
  {journal} {Proceedings of the IEEE}\ }\textbf {\bibinfo {volume} {108}},\
  \bibinfo {pages} {1684} (\bibinfo {year} {2020}{\natexlab{a}})}\BibitemShut
  {NoStop}%
\bibitem [{\citenamefont {Lifshitz}\ and\ \citenamefont
  {Pitaevskii}(1995)}]{MO-physics}%
  \BibitemOpen
  \bibfield  {author} {\bibinfo {author} {\bibfnamefont {E.}~\bibnamefont
  {Lifshitz}}\ and\ \bibinfo {author} {\bibfnamefont {L.}~\bibnamefont
  {Pitaevskii}},\ }\href {https://books.google.com.cu/books?id=B2jGpmL5SxEC}
  {\emph {\bibinfo {title} {Physical Kinetics: Volume 10}}},\ Course of
  theoretical physics\ (\bibinfo  {publisher} {Elsevier Science},\ \bibinfo
  {year} {1995})\BibitemShut {NoStop}%
\bibitem [{\citenamefont {Belotelov}\ \emph {et~al.}(2011)\citenamefont
  {Belotelov}, \citenamefont {Akimov}, \citenamefont {Pohl}, \citenamefont
  {Kotov}, \citenamefont {Kasture}, \citenamefont {Vengurlekar}, \citenamefont
  {Gopal}, \citenamefont {Yakovlev}, \citenamefont {Zvezdin},\ and\
  \citenamefont {Bayer}}]{MO-magnetophotoniccrstals}%
  \BibitemOpen
  \bibfield  {author} {\bibinfo {author} {\bibfnamefont {V.~I.}\ \bibnamefont
  {Belotelov}}, \bibinfo {author} {\bibfnamefont {I.~A.}\ \bibnamefont
  {Akimov}}, \bibinfo {author} {\bibfnamefont {M.}~\bibnamefont {Pohl}},
  \bibinfo {author} {\bibfnamefont {V.~A.}\ \bibnamefont {Kotov}}, \bibinfo
  {author} {\bibfnamefont {S.}~\bibnamefont {Kasture}}, \bibinfo {author}
  {\bibfnamefont {A.~S.}\ \bibnamefont {Vengurlekar}}, \bibinfo {author}
  {\bibfnamefont {A.~V.}\ \bibnamefont {Gopal}}, \bibinfo {author}
  {\bibfnamefont {D.~R.}\ \bibnamefont {Yakovlev}}, \bibinfo {author}
  {\bibfnamefont {A.~K.}\ \bibnamefont {Zvezdin}},\ and\ \bibinfo {author}
  {\bibfnamefont {M.}~\bibnamefont {Bayer}},\ }\bibfield  {title} {\bibinfo
  {title} {Enhanced magneto-optical effects in magnetoplasmonic crystals},\
  }\href {https://doi.org/10.1038/nnano.2011.54} {\bibfield  {journal}
  {\bibinfo  {journal} {Nature Nanotechnology}\ }\textbf {\bibinfo {volume}
  {6}},\ \bibinfo {pages} {370} (\bibinfo {year} {2011})}\BibitemShut {NoStop}%
\bibitem [{\citenamefont {Inoue}\ \emph {et~al.}(1999)\citenamefont {Inoue},
  \citenamefont {Arai}, \citenamefont {Fujii},\ and\ \citenamefont
  {Abe}}]{MO-PHC-1}%
  \BibitemOpen
  \bibfield  {author} {\bibinfo {author} {\bibfnamefont {M.}~\bibnamefont
  {Inoue}}, \bibinfo {author} {\bibfnamefont {K.~N.}\ \bibnamefont {Arai}},
  \bibinfo {author} {\bibfnamefont {T.}~\bibnamefont {Fujii}},\ and\ \bibinfo
  {author} {\bibfnamefont {M.}~\bibnamefont {Abe}},\ }\bibfield  {title}
  {\bibinfo {title} {One-dimensional magnetophotonic crystals},\ }\href
  {https://doi.org/10.1063/1.370120} {\bibfield  {journal} {\bibinfo  {journal}
  {Journal of Applied Physics}\ }\textbf {\bibinfo {volume} {85}},\ \bibinfo
  {pages} {5768} (\bibinfo {year} {1999})}\BibitemShut {NoStop}%
\bibitem [{\citenamefont {Takeda}\ \emph {et~al.}(2000)\citenamefont {Takeda},
  \citenamefont {Todoroki}, \citenamefont {Kitamoto}, \citenamefont {Abe},
  \citenamefont {Inoue}, \citenamefont {Fujii},\ and\ \citenamefont
  {Arai}}]{MO-PHC-2}%
  \BibitemOpen
  \bibfield  {author} {\bibinfo {author} {\bibfnamefont {E.}~\bibnamefont
  {Takeda}}, \bibinfo {author} {\bibfnamefont {N.}~\bibnamefont {Todoroki}},
  \bibinfo {author} {\bibfnamefont {Y.}~\bibnamefont {Kitamoto}}, \bibinfo
  {author} {\bibfnamefont {M.}~\bibnamefont {Abe}}, \bibinfo {author}
  {\bibfnamefont {M.}~\bibnamefont {Inoue}}, \bibinfo {author} {\bibfnamefont
  {T.}~\bibnamefont {Fujii}},\ and\ \bibinfo {author} {\bibfnamefont
  {K.}~\bibnamefont {Arai}},\ }\bibfield  {title} {\bibinfo {title} {Faraday
  effect enhancement in co-ferrite layer incorporated into one-dimensional
  photonic crystal working as a fabry-p{\'e}rot resonator},\ }\href
  {https://doi.org/10.1063/1.372840} {\bibfield  {journal} {\bibinfo  {journal}
  {Journal of Applied Physics}\ }\textbf {\bibinfo {volume} {87}},\ \bibinfo
  {pages} {6782} (\bibinfo {year} {2000})}\BibitemShut {NoStop}%
\bibitem [{\citenamefont {Steel}\ \emph {et~al.}(2000)\citenamefont {Steel},
  \citenamefont {Levy},\ and\ \citenamefont {Osgood}}]{MO-PHC-3}%
  \BibitemOpen
  \bibfield  {author} {\bibinfo {author} {\bibfnamefont {M.~J.}\ \bibnamefont
  {Steel}}, \bibinfo {author} {\bibfnamefont {M.}~\bibnamefont {Levy}},\ and\
  \bibinfo {author} {\bibfnamefont {R.~M.}\ \bibnamefont {Osgood}},\ }\bibfield
   {title} {\bibinfo {title} {High transmission enhanced faraday rotation in
  one-dimensional photonic crystals with defects},\ }\href
  {https://doi.org/10.1109/68.874225} {\bibfield  {journal} {\bibinfo
  {journal} {IEEE Photonics Technology Letters}\ }\textbf {\bibinfo {volume}
  {12}},\ \bibinfo {pages} {1171} (\bibinfo {year} {2000})}\BibitemShut
  {NoStop}%
\bibitem [{\citenamefont {Armelles}\ \emph {et~al.}(2013)\citenamefont
  {Armelles}, \citenamefont {Cebollada}, \citenamefont
  {Garc{\'i}a-Mart{\'i}n},\ and\ \citenamefont
  {Gonz{\'a}lez}}]{MO-magnetoplasmonics2}%
  \BibitemOpen
  \bibfield  {author} {\bibinfo {author} {\bibfnamefont {G.}~\bibnamefont
  {Armelles}}, \bibinfo {author} {\bibfnamefont {A.}~\bibnamefont {Cebollada}},
  \bibinfo {author} {\bibfnamefont {A.}~\bibnamefont {Garc{\'i}a-Mart{\'i}n}},\
  and\ \bibinfo {author} {\bibfnamefont {M.~U.}\ \bibnamefont {Gonz{\'a}lez}},\
  }\bibfield  {title} {\bibinfo {title} {Magnetoplasmonics: Combining magnetic
  and plasmonic functionalities},\ }\href
  {https://doi.org/10.1002/adom.201200011} {\bibfield  {journal} {\bibinfo
  {journal} {Advanced Optical Materials}\ }\textbf {\bibinfo {volume} {1}},\
  \bibinfo {pages} {10} (\bibinfo {year} {2013})}\BibitemShut {NoStop}%
\bibitem [{\citenamefont {Chin}\ \emph {et~al.}(2013)\citenamefont {Chin},
  \citenamefont {Steinle}, \citenamefont {Wehlus}, \citenamefont {Dregely},
  \citenamefont {Weiss}, \citenamefont {Belotelov}, \citenamefont {Stritzker},\
  and\ \citenamefont {Giessen}}]{MO-Plasmonics}%
  \BibitemOpen
  \bibfield  {author} {\bibinfo {author} {\bibfnamefont {J.~Y.}\ \bibnamefont
  {Chin}}, \bibinfo {author} {\bibfnamefont {T.}~\bibnamefont {Steinle}},
  \bibinfo {author} {\bibfnamefont {T.}~\bibnamefont {Wehlus}}, \bibinfo
  {author} {\bibfnamefont {D.}~\bibnamefont {Dregely}}, \bibinfo {author}
  {\bibfnamefont {T.}~\bibnamefont {Weiss}}, \bibinfo {author} {\bibfnamefont
  {V.~I.}\ \bibnamefont {Belotelov}}, \bibinfo {author} {\bibfnamefont
  {B.}~\bibnamefont {Stritzker}},\ and\ \bibinfo {author} {\bibfnamefont
  {H.}~\bibnamefont {Giessen}},\ }\bibfield  {title} {\bibinfo {title}
  {Nonreciprocal plasmonics enables giant enhancement of thin-film faraday
  rotation},\ }\bibfield  {journal} {\bibinfo  {journal} {Nature
  Communications}\ }\textbf {\bibinfo {volume} {4}},\ \href
  {https://doi.org/10.1038/ncomms2609} {10.1038/ncomms2609} (\bibinfo {year}
  {2013})\BibitemShut {NoStop}%
\bibitem [{\citenamefont {MacCaferri}\ \emph {et~al.}(2015)\citenamefont
  {MacCaferri}, \citenamefont {Inchausti}, \citenamefont
  {Garc{\'i}a-Mart{\'i}n}, \citenamefont {Cuevas}, \citenamefont {Tripathy},
  \citenamefont {Adeyeye},\ and\ \citenamefont {Vavassori}}]{MO-SPPs}%
  \BibitemOpen
  \bibfield  {author} {\bibinfo {author} {\bibfnamefont {N.}~\bibnamefont
  {MacCaferri}}, \bibinfo {author} {\bibfnamefont {X.}~\bibnamefont
  {Inchausti}}, \bibinfo {author} {\bibfnamefont {A.}~\bibnamefont
  {Garc{\'i}a-Mart{\'i}n}}, \bibinfo {author} {\bibfnamefont {J.~C.}\
  \bibnamefont {Cuevas}}, \bibinfo {author} {\bibfnamefont {D.}~\bibnamefont
  {Tripathy}}, \bibinfo {author} {\bibfnamefont {A.~O.}\ \bibnamefont
  {Adeyeye}},\ and\ \bibinfo {author} {\bibfnamefont {P.}~\bibnamefont
  {Vavassori}},\ }\bibfield  {title} {\bibinfo {title} {Resonant enhancement of
  magneto-optical activity induced by surface plasmon polariton modes coupling
  in 2d magnetoplasmonic crystals},\ }\href
  {https://doi.org/10.1021/acsphotonics.5b00490} {\bibfield  {journal}
  {\bibinfo  {journal} {ACS Photonics}\ }\textbf {\bibinfo {volume} {2}},\
  \bibinfo {pages} {1769} (\bibinfo {year} {2015})}\BibitemShut {NoStop}%
\bibitem [{\citenamefont {Asadchy}\ \emph
  {et~al.}(2020{\natexlab{b}})\citenamefont {Asadchy}, \citenamefont {Guo},
  \citenamefont {Zhao},\ and\ \citenamefont {Fan}}]{asadchy_sub_2020}%
  \BibitemOpen
  \bibfield  {author} {\bibinfo {author} {\bibfnamefont {V.~S.}\ \bibnamefont
  {Asadchy}}, \bibinfo {author} {\bibfnamefont {C.}~\bibnamefont {Guo}},
  \bibinfo {author} {\bibfnamefont {B.}~\bibnamefont {Zhao}},\ and\ \bibinfo
  {author} {\bibfnamefont {S.}~\bibnamefont {Fan}},\ }\bibfield  {title}
  {\bibinfo {title} {Sub-wavelength passive optical isolators using photonic
  structures based on {{Weyl}} semimetals},\ }\href
  {https://doi.org/10.1002/adom.202000100} {\bibfield  {journal} {\bibinfo
  {journal} {Advanced Optical Materials}\ }\textbf {\bibinfo {volume} {8}},\
  \bibinfo {pages} {2000100} (\bibinfo {year}
  {2020}{\natexlab{b}})}\BibitemShut {NoStop}%
\bibitem [{\citenamefont {Khaghani}\ \emph {et~al.}(2022)\citenamefont
  {Khaghani}, \citenamefont {Farzad},\ and\ \citenamefont {Asgari}}]{MO-layer}%
  \BibitemOpen
  \bibfield  {author} {\bibinfo {author} {\bibfnamefont {Z.}~\bibnamefont
  {Khaghani}}, \bibinfo {author} {\bibfnamefont {M.~H.}\ \bibnamefont
  {Farzad}},\ and\ \bibinfo {author} {\bibfnamefont {A.}~\bibnamefont
  {Asgari}},\ }\bibfield  {title} {\bibinfo {title} {Enhanced magneto-optical
  effect in three layer based magnetoplasmonic structures},\ }\bibfield
  {journal} {\bibinfo  {journal} {Optical and Quantum Electronics}\ }\textbf
  {\bibinfo {volume} {54}},\ \href {https://doi.org/10.1007/s11082-022-04012-z}
  {10.1007/s11082-022-04012-z} (\bibinfo {year} {2022})\BibitemShut {NoStop}%
\bibitem [{\citenamefont {Christofi}\ \emph {et~al.}(2018)\citenamefont
  {Christofi}, \citenamefont {Kawaguchi}, \citenamefont {Al{\`u}},\ and\
  \citenamefont {Khanikaev}}]{MO-Huygens-FOM}%
  \BibitemOpen
  \bibfield  {author} {\bibinfo {author} {\bibfnamefont {A.}~\bibnamefont
  {Christofi}}, \bibinfo {author} {\bibfnamefont {Y.}~\bibnamefont
  {Kawaguchi}}, \bibinfo {author} {\bibfnamefont {A.}~\bibnamefont {Al{\`u}}},\
  and\ \bibinfo {author} {\bibfnamefont {A.~B.}\ \bibnamefont {Khanikaev}},\
  }\bibfield  {title} {\bibinfo {title} {Giant enhancement of faraday rotation
  due to electromagnetically induced transparency in all-dielectric
  magneto-optical metasurfaces},\ }\href {https://doi.org/10.1364/ol.43.001838}
  {\bibfield  {journal} {\bibinfo  {journal} {Optics Letters}\ }\textbf
  {\bibinfo {volume} {43}},\ \bibinfo {pages} {1838} (\bibinfo {year}
  {2018})}\BibitemShut {NoStop}%
\bibitem [{\citenamefont {Xia}\ \emph {et~al.}(2022)\citenamefont {Xia},
  \citenamefont {Ignatyeva}, \citenamefont {Liu}, \citenamefont {Qin},
  \citenamefont {Kang}, \citenamefont {Yang}, \citenamefont {Chen},
  \citenamefont {Duan}, \citenamefont {Deng}, \citenamefont {Long},
  \citenamefont {Veis}, \citenamefont {Belotelov},\ and\ \citenamefont
  {Bi}}]{MO-meta-FOM}%
  \BibitemOpen
  \bibfield  {author} {\bibinfo {author} {\bibfnamefont {S.}~\bibnamefont
  {Xia}}, \bibinfo {author} {\bibfnamefont {D.~O.}\ \bibnamefont {Ignatyeva}},
  \bibinfo {author} {\bibfnamefont {Q.}~\bibnamefont {Liu}}, \bibinfo {author}
  {\bibfnamefont {J.}~\bibnamefont {Qin}}, \bibinfo {author} {\bibfnamefont
  {T.}~\bibnamefont {Kang}}, \bibinfo {author} {\bibfnamefont {W.}~\bibnamefont
  {Yang}}, \bibinfo {author} {\bibfnamefont {Y.}~\bibnamefont {Chen}}, \bibinfo
  {author} {\bibfnamefont {H.}~\bibnamefont {Duan}}, \bibinfo {author}
  {\bibfnamefont {L.}~\bibnamefont {Deng}}, \bibinfo {author} {\bibfnamefont
  {D.}~\bibnamefont {Long}}, \bibinfo {author} {\bibfnamefont {M.}~\bibnamefont
  {Veis}}, \bibinfo {author} {\bibfnamefont {V.~I.}\ \bibnamefont
  {Belotelov}},\ and\ \bibinfo {author} {\bibfnamefont {L.}~\bibnamefont
  {Bi}},\ }\bibfield  {title} {\bibinfo {title} {Enhancement of the faraday
  effect and magneto-optical figure of merit in all-dielectric metasurfaces},\
  }\href {https://doi.org/10.1021/acsphotonics.1c01692} {\bibfield  {journal}
  {\bibinfo  {journal} {ACS Photonics}\ }\textbf {\bibinfo {volume} {9}},\
  \bibinfo {pages} {1240} (\bibinfo {year} {2022})}\BibitemShut {NoStop}%
\bibitem [{\citenamefont {Barsukova}\ \emph {et~al.}(2019)\citenamefont
  {Barsukova}, \citenamefont {Musorin}, \citenamefont {Shorokhov},\ and\
  \citenamefont {Fedyanin}}]{MO-hybrid}%
  \BibitemOpen
  \bibfield  {author} {\bibinfo {author} {\bibfnamefont {M.~G.}\ \bibnamefont
  {Barsukova}}, \bibinfo {author} {\bibfnamefont {A.~I.}\ \bibnamefont
  {Musorin}}, \bibinfo {author} {\bibfnamefont {A.~S.}\ \bibnamefont
  {Shorokhov}},\ and\ \bibinfo {author} {\bibfnamefont {A.~A.}\ \bibnamefont
  {Fedyanin}},\ }\bibfield  {title} {\bibinfo {title} {Enhanced magneto-optical
  effects in hybrid ni-si metasurfaces},\ }\bibfield  {journal} {\bibinfo
  {journal} {APL Photonics}\ }\textbf {\bibinfo {volume} {4}},\ \href
  {https://doi.org/10.1063/1.5066307} {10.1063/1.5066307} (\bibinfo {year}
  {2019})\BibitemShut {NoStop}%
\bibitem [{\citenamefont {Ignatyeva}\ \emph {et~al.}(2020)\citenamefont
  {Ignatyeva}, \citenamefont {Karki}, \citenamefont {Voronov}, \citenamefont
  {Kozhaev}, \citenamefont {Krichevsky}, \citenamefont {Chernov}, \citenamefont
  {Levy},\ and\ \citenamefont {Belotelov}}]{MO-BICs-3}%
  \BibitemOpen
  \bibfield  {author} {\bibinfo {author} {\bibfnamefont {D.~O.}\ \bibnamefont
  {Ignatyeva}}, \bibinfo {author} {\bibfnamefont {D.}~\bibnamefont {Karki}},
  \bibinfo {author} {\bibfnamefont {A.~A.}\ \bibnamefont {Voronov}}, \bibinfo
  {author} {\bibfnamefont {M.~A.}\ \bibnamefont {Kozhaev}}, \bibinfo {author}
  {\bibfnamefont {D.~M.}\ \bibnamefont {Krichevsky}}, \bibinfo {author}
  {\bibfnamefont {A.~I.}\ \bibnamefont {Chernov}}, \bibinfo {author}
  {\bibfnamefont {M.}~\bibnamefont {Levy}},\ and\ \bibinfo {author}
  {\bibfnamefont {V.~I.}\ \bibnamefont {Belotelov}},\ }\bibfield  {title}
  {\bibinfo {title} {All-dielectric magnetic metasurface for advanced light
  control in dual polarizations combined with high-q resonances},\ }\bibfield
  {journal} {\bibinfo  {journal} {Nature Communications}\ }\textbf {\bibinfo
  {volume} {11}},\ \href {https://doi.org/10.1038/s41467-020-19310-x}
  {10.1038/s41467-020-19310-x} (\bibinfo {year} {2020})\BibitemShut {NoStop}%
\bibitem [{\citenamefont {Kupriianov}\ \emph {et~al.}(2019)\citenamefont
  {Kupriianov}, \citenamefont {Xu}, \citenamefont {Sayanskiy}, \citenamefont
  {Dmitriev}, \citenamefont {Kivshar},\ and\ \citenamefont {Tuz}}]{Meta-BICs}%
  \BibitemOpen
  \bibfield  {author} {\bibinfo {author} {\bibfnamefont {A.~S.}\ \bibnamefont
  {Kupriianov}}, \bibinfo {author} {\bibfnamefont {Y.}~\bibnamefont {Xu}},
  \bibinfo {author} {\bibfnamefont {A.}~\bibnamefont {Sayanskiy}}, \bibinfo
  {author} {\bibfnamefont {V.}~\bibnamefont {Dmitriev}}, \bibinfo {author}
  {\bibfnamefont {Y.~S.}\ \bibnamefont {Kivshar}},\ and\ \bibinfo {author}
  {\bibfnamefont {V.~R.}\ \bibnamefont {Tuz}},\ }\bibfield  {title} {\bibinfo
  {title} {Metasurface engineering through bound states in the continuum},\
  }\bibfield  {journal} {\bibinfo  {journal} {Physical Review Applied}\
  }\textbf {\bibinfo {volume} {12}},\ \href
  {https://doi.org/10.1103/PhysRevApplied.12.014024}
  {10.1103/PhysRevApplied.12.014024} (\bibinfo {year} {2019})\BibitemShut
  {NoStop}%
\bibitem [{\citenamefont {Hsu}\ \emph {et~al.}(2016)\citenamefont {Hsu},
  \citenamefont {Zhen}, \citenamefont {Stone}, \citenamefont {Joannopoulos},\
  and\ \citenamefont {Soljacic}}]{BICs-review}%
  \BibitemOpen
  \bibfield  {author} {\bibinfo {author} {\bibfnamefont {C.~W.}\ \bibnamefont
  {Hsu}}, \bibinfo {author} {\bibfnamefont {B.}~\bibnamefont {Zhen}}, \bibinfo
  {author} {\bibfnamefont {A.~D.}\ \bibnamefont {Stone}}, \bibinfo {author}
  {\bibfnamefont {J.~D.}\ \bibnamefont {Joannopoulos}},\ and\ \bibinfo {author}
  {\bibfnamefont {M.}~\bibnamefont {Soljacic}},\ }\bibfield  {title} {\bibinfo
  {title} {Bound states in the continuum},\ }\bibfield  {journal} {\bibinfo
  {journal} {Nature Reviews Materials}\ }\textbf {\bibinfo {volume} {1}},\
  \href {https://doi.org/10.1038/natrevmats.2016.48}
  {10.1038/natrevmats.2016.48} (\bibinfo {year} {2016})\BibitemShut {NoStop}%
\bibitem [{\citenamefont {He}\ \emph {et~al.}(2018)\citenamefont {He},
  \citenamefont {Guo}, \citenamefont {Feng}, \citenamefont {Xu},\ and\
  \citenamefont {Miroshnichenko}}]{BIC-1}%
  \BibitemOpen
  \bibfield  {author} {\bibinfo {author} {\bibfnamefont {Y.}~\bibnamefont
  {He}}, \bibinfo {author} {\bibfnamefont {G.}~\bibnamefont {Guo}}, \bibinfo
  {author} {\bibfnamefont {T.}~\bibnamefont {Feng}}, \bibinfo {author}
  {\bibfnamefont {Y.}~\bibnamefont {Xu}},\ and\ \bibinfo {author}
  {\bibfnamefont {A.~E.}\ \bibnamefont {Miroshnichenko}},\ }\bibfield  {title}
  {\bibinfo {title} {Toroidal dipole bound states in the continuum},\
  }\bibfield  {journal} {\bibinfo  {journal} {Physical Review B}\ }\textbf
  {\bibinfo {volume} {98}},\ \href {https://doi.org/10.1103/PhysRevB.98.161112}
  {10.1103/PhysRevB.98.161112} (\bibinfo {year} {2018})\BibitemShut {NoStop}%
\bibitem [{\citenamefont {Lee}\ \emph {et~al.}(2012)\citenamefont {Lee},
  \citenamefont {Zhen}, \citenamefont {Chua}, \citenamefont {Qiu},
  \citenamefont {Joannopoulos}, \citenamefont {Solja{\^c}i{\'c}},\ and\
  \citenamefont {Shapira}}]{BIC-2}%
  \BibitemOpen
  \bibfield  {author} {\bibinfo {author} {\bibfnamefont {J.}~\bibnamefont
  {Lee}}, \bibinfo {author} {\bibfnamefont {B.}~\bibnamefont {Zhen}}, \bibinfo
  {author} {\bibfnamefont {S.~L.}\ \bibnamefont {Chua}}, \bibinfo {author}
  {\bibfnamefont {W.}~\bibnamefont {Qiu}}, \bibinfo {author} {\bibfnamefont
  {J.~D.}\ \bibnamefont {Joannopoulos}}, \bibinfo {author} {\bibfnamefont
  {M.}~\bibnamefont {Solja{\^c}i{\'c}}},\ and\ \bibinfo {author} {\bibfnamefont
  {O.}~\bibnamefont {Shapira}},\ }\bibfield  {title} {\bibinfo {title}
  {Observation and differentiation of unique high-q optical resonances near
  zero wave vector in macroscopic photonic crystal slabs},\ }\bibfield
  {journal} {\bibinfo  {journal} {Physical Review Letters}\ }\textbf {\bibinfo
  {volume} {109}},\ \href {https://doi.org/10.1103/PhysRevLett.109.067401}
  {10.1103/PhysRevLett.109.067401} (\bibinfo {year} {2012})\BibitemShut
  {NoStop}%
\bibitem [{\citenamefont {Tang}\ \emph {et~al.}(2023)\citenamefont {Tang},
  \citenamefont {Liang}, \citenamefont {Yao}, \citenamefont {Chen},
  \citenamefont {Lin}, \citenamefont {Wang}, \citenamefont {Zhang},
  \citenamefont {Huang}, \citenamefont {Yu},\ and\ \citenamefont
  {Tsai}}]{BIC-CD}%
  \BibitemOpen
  \bibfield  {author} {\bibinfo {author} {\bibfnamefont {Y.}~\bibnamefont
  {Tang}}, \bibinfo {author} {\bibfnamefont {Y.}~\bibnamefont {Liang}},
  \bibinfo {author} {\bibfnamefont {J.}~\bibnamefont {Yao}}, \bibinfo {author}
  {\bibfnamefont {M.~K.}\ \bibnamefont {Chen}}, \bibinfo {author}
  {\bibfnamefont {S.}~\bibnamefont {Lin}}, \bibinfo {author} {\bibfnamefont
  {Z.}~\bibnamefont {Wang}}, \bibinfo {author} {\bibfnamefont {J.}~\bibnamefont
  {Zhang}}, \bibinfo {author} {\bibfnamefont {X.~G.}\ \bibnamefont {Huang}},
  \bibinfo {author} {\bibfnamefont {C.}~\bibnamefont {Yu}},\ and\ \bibinfo
  {author} {\bibfnamefont {D.~P.}\ \bibnamefont {Tsai}},\ }\bibfield  {title}
  {\bibinfo {title} {Chiral bound states in the continuum in plasmonic
  metasurfaces},\ }\bibfield  {journal} {\bibinfo  {journal} {Laser and
  Photonics Reviews}\ }\href {https://doi.org/10.1002/lpor.202200597}
  {10.1002/lpor.202200597} (\bibinfo {year} {2023})\BibitemShut {NoStop}%
\bibitem [{\citenamefont {Gorkunov}\ \emph {et~al.}(2020)\citenamefont
  {Gorkunov}, \citenamefont {Antonov},\ and\ \citenamefont
  {Kivshar}}]{Maximun-chirality-kivshar}%
  \BibitemOpen
  \bibfield  {author} {\bibinfo {author} {\bibfnamefont {M.~V.}\ \bibnamefont
  {Gorkunov}}, \bibinfo {author} {\bibfnamefont {A.~A.}\ \bibnamefont
  {Antonov}},\ and\ \bibinfo {author} {\bibfnamefont {Y.~S.}\ \bibnamefont
  {Kivshar}},\ }\bibfield  {title} {\bibinfo {title} {Metasurfaces with maximum
  chirality empowered by bound states in the continuum},\ }\bibfield  {journal}
  {\bibinfo  {journal} {Physical Review Letters}\ }\textbf {\bibinfo {volume}
  {125}},\ \href {https://doi.org/10.1103/PhysRevLett.125.093903}
  {10.1103/PhysRevLett.125.093903} (\bibinfo {year} {2020})\BibitemShut
  {NoStop}%
\bibitem [{\citenamefont {Gorkunov}\ \emph {et~al.}(2021)\citenamefont
  {Gorkunov}, \citenamefont {Antonov}, \citenamefont {Tuz}, \citenamefont
  {Kupriianov},\ and\ \citenamefont {Kivshar}}]{BICs-chirality}%
  \BibitemOpen
  \bibfield  {author} {\bibinfo {author} {\bibfnamefont {M.~V.}\ \bibnamefont
  {Gorkunov}}, \bibinfo {author} {\bibfnamefont {A.~A.}\ \bibnamefont
  {Antonov}}, \bibinfo {author} {\bibfnamefont {V.~R.}\ \bibnamefont {Tuz}},
  \bibinfo {author} {\bibfnamefont {A.~S.}\ \bibnamefont {Kupriianov}},\ and\
  \bibinfo {author} {\bibfnamefont {Y.~S.}\ \bibnamefont {Kivshar}},\
  }\bibfield  {title} {\bibinfo {title} {Bound states in the continuum underpin
  near-lossless maximum chirality in dielectric metasurfaces},\ }\bibfield
  {journal} {\bibinfo  {journal} {Advanced Optical Materials}\ }\textbf
  {\bibinfo {volume} {9}},\ \href {https://doi.org/10.1002/adom.202100797}
  {10.1002/adom.202100797} (\bibinfo {year} {2021})\BibitemShut {NoStop}%
\bibitem [{\citenamefont {Zong}\ \emph {et~al.}(2023)\citenamefont {Zong},
  \citenamefont {Li},\ and\ \citenamefont {Liu}}]{BIC-absorption}%
  \BibitemOpen
  \bibfield  {author} {\bibinfo {author} {\bibfnamefont {X.}~\bibnamefont
  {Zong}}, \bibinfo {author} {\bibfnamefont {L.}~\bibnamefont {Li}},\ and\
  \bibinfo {author} {\bibfnamefont {Y.}~\bibnamefont {Liu}},\ }\bibfield
  {title} {\bibinfo {title} {Bound states in the continuum enabling
  ultra-narrowband perfect absorption},\ }\bibfield  {journal} {\bibinfo
  {journal} {New Journal of Physics}\ }\textbf {\bibinfo {volume} {25}},\ \href
  {https://doi.org/10.1088/1367-2630/acb9b3} {10.1088/1367-2630/acb9b3}
  (\bibinfo {year} {2023})\BibitemShut {NoStop}%
\bibitem [{\citenamefont {Tian}\ \emph {et~al.}(2020)\citenamefont {Tian},
  \citenamefont {Li}, \citenamefont {Belov}, \citenamefont {Sinha},
  \citenamefont {Qian},\ and\ \citenamefont {Qiu}}]{BIC-absorption2}%
  \BibitemOpen
  \bibfield  {author} {\bibinfo {author} {\bibfnamefont {J.}~\bibnamefont
  {Tian}}, \bibinfo {author} {\bibfnamefont {Q.}~\bibnamefont {Li}}, \bibinfo
  {author} {\bibfnamefont {P.~A.}\ \bibnamefont {Belov}}, \bibinfo {author}
  {\bibfnamefont {R.~K.}\ \bibnamefont {Sinha}}, \bibinfo {author}
  {\bibfnamefont {W.}~\bibnamefont {Qian}},\ and\ \bibinfo {author}
  {\bibfnamefont {M.}~\bibnamefont {Qiu}},\ }\bibfield  {title} {\bibinfo
  {title} {High- q all-dielectric metasurface: Super and suppressed optical
  absorption},\ }\href {https://doi.org/10.1021/acsphotonics.0c00003}
  {\bibfield  {journal} {\bibinfo  {journal} {ACS Photonics}\ }\textbf
  {\bibinfo {volume} {7}},\ \bibinfo {pages} {1436} (\bibinfo {year}
  {2020})}\BibitemShut {NoStop}%
\bibitem [{\citenamefont {Koshelev}\ \emph {et~al.}(2018)\citenamefont
  {Koshelev}, \citenamefont {Lepeshov}, \citenamefont {Liu}, \citenamefont
  {Bogdanov},\ and\ \citenamefont {Kivshar}}]{BICs-SP}%
  \BibitemOpen
  \bibfield  {author} {\bibinfo {author} {\bibfnamefont {K.}~\bibnamefont
  {Koshelev}}, \bibinfo {author} {\bibfnamefont {S.}~\bibnamefont {Lepeshov}},
  \bibinfo {author} {\bibfnamefont {M.}~\bibnamefont {Liu}}, \bibinfo {author}
  {\bibfnamefont {A.}~\bibnamefont {Bogdanov}},\ and\ \bibinfo {author}
  {\bibfnamefont {Y.}~\bibnamefont {Kivshar}},\ }\bibfield  {title} {\bibinfo
  {title} {Asymmetric metasurfaces with high- q resonances governed by bound
  states in the continuum},\ }\bibfield  {journal} {\bibinfo  {journal}
  {Physical Review Letters}\ }\textbf {\bibinfo {volume} {121}},\ \href
  {https://doi.org/10.1103/PhysRevLett.121.193903}
  {10.1103/PhysRevLett.121.193903} (\bibinfo {year} {2018})\BibitemShut
  {NoStop}%
\bibitem [{\citenamefont {Yu}\ \emph {et~al.}(2018)\citenamefont {Yu},
  \citenamefont {Kupriianov}, \citenamefont {Dmitriev},\ and\ \citenamefont
  {Tuz}}]{BICs-grupos}%
  \BibitemOpen
  \bibfield  {author} {\bibinfo {author} {\bibfnamefont {P.}~\bibnamefont
  {Yu}}, \bibinfo {author} {\bibfnamefont {A.~S.}\ \bibnamefont {Kupriianov}},
  \bibinfo {author} {\bibfnamefont {V.}~\bibnamefont {Dmitriev}},\ and\
  \bibinfo {author} {\bibfnamefont {V.~R.}\ \bibnamefont {Tuz}},\ }\bibfield
  {title} {\bibinfo {title} {All-dielectric metasurfaces with trapped modes:
  group-theoretical description},\ }\bibfield  {journal} {\bibinfo  {journal}
  {Journal of Applied Physics}\ }\href {https://doi.org/10.1063/1.5087054}
  {10.1063/1.5087054} (\bibinfo {year} {2018})\BibitemShut {NoStop}%
\bibitem [{\citenamefont {Sakoda}(1995)}]{BICs-SP-2}%
  \BibitemOpen
  \bibfield  {author} {\bibinfo {author} {\bibfnamefont {K.}~\bibnamefont
  {Sakoda}},\ }\bibfield  {title} {\bibinfo {title} {Symmetry, degeneracy, and
  uncoupled modes in two-dimensional photonic lattices},\ }\href
  {https://doi.org/10.1103/PhysRevB.52.7982} {\bibfield  {journal} {\bibinfo
  {journal} {Phys. Rev. B}\ }\textbf {\bibinfo {volume} {52}},\ \bibinfo
  {pages} {7982} (\bibinfo {year} {1995})}\BibitemShut {NoStop}%
\bibitem [{\citenamefont {Overvig}\ \emph {et~al.}(2020)\citenamefont
  {Overvig}, \citenamefont {Malek}, \citenamefont {Carter}, \citenamefont
  {Shrestha},\ and\ \citenamefont {Yu}}]{BICs-SP-3}%
  \BibitemOpen
  \bibfield  {author} {\bibinfo {author} {\bibfnamefont {A.~C.}\ \bibnamefont
  {Overvig}}, \bibinfo {author} {\bibfnamefont {S.~C.}\ \bibnamefont {Malek}},
  \bibinfo {author} {\bibfnamefont {M.~J.}\ \bibnamefont {Carter}}, \bibinfo
  {author} {\bibfnamefont {S.}~\bibnamefont {Shrestha}},\ and\ \bibinfo
  {author} {\bibfnamefont {N.}~\bibnamefont {Yu}},\ }\bibfield  {title}
  {\bibinfo {title} {Selection rules for quasibound states in the continuum},\
  }\href {https://doi.org/10.1103/PhysRevB.102.035434} {\bibfield  {journal}
  {\bibinfo  {journal} {Phys. Rev. B}\ }\textbf {\bibinfo {volume} {102}},\
  \bibinfo {pages} {035434} (\bibinfo {year} {2020})}\BibitemShut {NoStop}%
\bibitem [{\citenamefont {Chen}\ \emph {et~al.}(2019)\citenamefont {Chen},
  \citenamefont {Zhang},\ and\ \citenamefont {Zhang}}]{MO-BICs-2}%
  \BibitemOpen
  \bibfield  {author} {\bibinfo {author} {\bibfnamefont {G.~Y.}\ \bibnamefont
  {Chen}}, \bibinfo {author} {\bibfnamefont {W.~X.}\ \bibnamefont {Zhang}},\
  and\ \bibinfo {author} {\bibfnamefont {X.~D.}\ \bibnamefont {Zhang}},\
  }\bibfield  {title} {\bibinfo {title} {Strong terahertz magneto-optical
  phenomena based on quasi-bound states in the continuum and fano resonances},\
  }\href {https://doi.org/10.1364/oe.27.016449} {\bibfield  {journal} {\bibinfo
   {journal} {Optics Express}\ }\textbf {\bibinfo {volume} {27}},\ \bibinfo
  {pages} {16449} (\bibinfo {year} {2019})}\BibitemShut {NoStop}%
\bibitem [{\citenamefont {Chernyak}\ \emph {et~al.}(2020)\citenamefont
  {Chernyak}, \citenamefont {Barsukova}, \citenamefont {Shorokhov},
  \citenamefont {Musorin},\ and\ \citenamefont {Fedyanin}}]{MO-BICs}%
  \BibitemOpen
  \bibfield  {author} {\bibinfo {author} {\bibfnamefont {A.~M.}\ \bibnamefont
  {Chernyak}}, \bibinfo {author} {\bibfnamefont {M.~G.}\ \bibnamefont
  {Barsukova}}, \bibinfo {author} {\bibfnamefont {A.~S.}\ \bibnamefont
  {Shorokhov}}, \bibinfo {author} {\bibfnamefont {A.~I.}\ \bibnamefont
  {Musorin}},\ and\ \bibinfo {author} {\bibfnamefont {A.~A.}\ \bibnamefont
  {Fedyanin}},\ }\bibfield  {title} {\bibinfo {title} {Bound states in the
  continuum in magnetophotonic metasurfaces},\ }\href
  {https://doi.org/10.1134/S0021364020010105} {\bibfield  {journal} {\bibinfo
  {journal} {JETP Letters}\ }\textbf {\bibinfo {volume} {111}},\ \bibinfo
  {pages} {46} (\bibinfo {year} {2020})}\BibitemShut {NoStop}%
\bibitem [{\citenamefont {Abujetas}\ \emph {et~al.}(2021)\citenamefont
  {Abujetas}, \citenamefont {de~Sousa}, \citenamefont
  {{Garc{\'i}a-Mart{\'i}n}}, \citenamefont {Llorens},\ and\ \citenamefont
  {{S{\'a}nchez-Gil}}}]{abujetas_active_2021}%
  \BibitemOpen
  \bibfield  {author} {\bibinfo {author} {\bibfnamefont {D.~R.}\ \bibnamefont
  {Abujetas}}, \bibinfo {author} {\bibfnamefont {N.}~\bibnamefont {de~Sousa}},
  \bibinfo {author} {\bibfnamefont {A.}~\bibnamefont
  {{Garc{\'i}a-Mart{\'i}n}}}, \bibinfo {author} {\bibfnamefont {J.~M.}\
  \bibnamefont {Llorens}},\ and\ \bibinfo {author} {\bibfnamefont {J.~A.}\
  \bibnamefont {{S{\'a}nchez-Gil}}},\ }\bibfield  {title} {\bibinfo {title}
  {Active angular tuning and switching of {{Brewster}} quasi bound states in
  the continuum in magneto-optic metasurfaces},\ }\href
  {https://doi.org/10.1515/nanoph-2021-0412} {\bibfield  {journal} {\bibinfo
  {journal} {Nanophotonics}\ }\textbf {\bibinfo {volume} {10}},\ \bibinfo
  {pages} {4223} (\bibinfo {year} {2021})}\BibitemShut {NoStop}%
\bibitem [{\citenamefont {Suh}\ \emph {et~al.}(2004)\citenamefont {Suh},
  \citenamefont {Wang},\ and\ \citenamefont {Fan}}]{CMT-OG}%
  \BibitemOpen
  \bibfield  {author} {\bibinfo {author} {\bibfnamefont {W.}~\bibnamefont
  {Suh}}, \bibinfo {author} {\bibfnamefont {Z.}~\bibnamefont {Wang}},\ and\
  \bibinfo {author} {\bibfnamefont {S.}~\bibnamefont {Fan}},\ }\bibfield
  {title} {\bibinfo {title} {Temporal coupled-mode theory and the presence of
  non-orthogonal modes in lossless multimode cavities},\ }\href
  {https://doi.org/10.1109/JQE.2004.834773} {\bibfield  {journal} {\bibinfo
  {journal} {IEEE Journal of Quantum Electronics}\ }\textbf {\bibinfo {volume}
  {40}},\ \bibinfo {pages} {1511} (\bibinfo {year} {2004})}\BibitemShut
  {NoStop}%
\bibitem [{\citenamefont {Zhao}\ \emph {et~al.}(2019)\citenamefont {Zhao},
  \citenamefont {Guo},\ and\ \citenamefont {Fan}}]{CMT-eqs}%
  \BibitemOpen
  \bibfield  {author} {\bibinfo {author} {\bibfnamefont {Z.}~\bibnamefont
  {Zhao}}, \bibinfo {author} {\bibfnamefont {C.}~\bibnamefont {Guo}},\ and\
  \bibinfo {author} {\bibfnamefont {S.}~\bibnamefont {Fan}},\ }\bibfield
  {title} {\bibinfo {title} {Connection of temporal coupled-mode-theory
  formalisms for a resonant optical system and its time-reversal conjugate},\
  }\bibfield  {journal} {\bibinfo  {journal} {Physical Review A}\ }\textbf
  {\bibinfo {volume} {99}},\ \href {https://doi.org/10.1103/PhysRevA.99.033839}
  {10.1103/PhysRevA.99.033839} (\bibinfo {year} {2019})\BibitemShut {NoStop}%
\bibitem [{\citenamefont {Chen}\ \emph
  {et~al.}(2018{\natexlab{a}})\citenamefont {Chen}, \citenamefont {Kim},
  \citenamefont {Wong},\ and\ \citenamefont {Eleftheriades}}]{Huygens-pairs}%
  \BibitemOpen
  \bibfield  {author} {\bibinfo {author} {\bibfnamefont {M.}~\bibnamefont
  {Chen}}, \bibinfo {author} {\bibfnamefont {M.}~\bibnamefont {Kim}}, \bibinfo
  {author} {\bibfnamefont {A.~M.}\ \bibnamefont {Wong}},\ and\ \bibinfo
  {author} {\bibfnamefont {G.~V.}\ \bibnamefont {Eleftheriades}},\ }\bibfield
  {title} {\bibinfo {title} {Huygens' metasurfaces from microwaves to optics: a
  review},\ }\href {https://doi.org/doi:10.1515/nanoph-2017-0117} {\bibfield
  {journal} {\bibinfo  {journal} {Nanophotonics}\ }\textbf {\bibinfo {volume}
  {7}},\ \bibinfo {pages} {1207} (\bibinfo {year}
  {2018}{\natexlab{a}})}\BibitemShut {NoStop}%
\bibitem [{\citenamefont {Lawrence}\ and\ \citenamefont
  {Dionne}(2019)}]{Geometry-1}%
  \BibitemOpen
  \bibfield  {author} {\bibinfo {author} {\bibfnamefont {M.}~\bibnamefont
  {Lawrence}}\ and\ \bibinfo {author} {\bibfnamefont {J.~A.}\ \bibnamefont
  {Dionne}},\ }\bibfield  {title} {\bibinfo {title} {Nanoscale nonreciprocity
  via photon-spin-polarized stimulated raman scattering},\ }\bibfield
  {journal} {\bibinfo  {journal} {Nature Communications}\ }\textbf {\bibinfo
  {volume} {10}},\ \href {https://doi.org/10.1038/s41467-019-11175-z}
  {10.1038/s41467-019-11175-z} (\bibinfo {year} {2019})\BibitemShut {NoStop}%
\bibitem [{\citenamefont {Hu}\ \emph {et~al.}(2020)\citenamefont {Hu},
  \citenamefont {Lawrence},\ and\ \citenamefont {Dionne}}]{Geometry-2}%
  \BibitemOpen
  \bibfield  {author} {\bibinfo {author} {\bibfnamefont {J.}~\bibnamefont
  {Hu}}, \bibinfo {author} {\bibfnamefont {M.}~\bibnamefont {Lawrence}},\ and\
  \bibinfo {author} {\bibfnamefont {J.~A.}\ \bibnamefont {Dionne}},\ }\bibfield
   {title} {\bibinfo {title} {High quality factor dielectric metasurfaces for
  ultraviolet circular dichroism spectroscopy},\ }\href
  {https://doi.org/10.1021/acsphotonics.9b01352} {\bibfield  {journal}
  {\bibinfo  {journal} {ACS Photonics}\ }\textbf {\bibinfo {volume} {7}},\
  \bibinfo {pages} {36} (\bibinfo {year} {2020})}\BibitemShut {NoStop}%
\bibitem [{\citenamefont {Jesenska}\ \emph {et~al.}(2016)\citenamefont
  {Jesenska}, \citenamefont {Yoshida}, \citenamefont {Shinozaki}, \citenamefont
  {Ishibashi}, \citenamefont {Beran}, \citenamefont {Zahradnik}, \citenamefont
  {Antos}, \citenamefont {Ku{\^c}era},\ and\ \citenamefont
  {Veis}}]{Meta-BiYIG-material}%
  \BibitemOpen
  \bibfield  {author} {\bibinfo {author} {\bibfnamefont {E.}~\bibnamefont
  {Jesenska}}, \bibinfo {author} {\bibfnamefont {T.}~\bibnamefont {Yoshida}},
  \bibinfo {author} {\bibfnamefont {K.}~\bibnamefont {Shinozaki}}, \bibinfo
  {author} {\bibfnamefont {T.}~\bibnamefont {Ishibashi}}, \bibinfo {author}
  {\bibfnamefont {L.}~\bibnamefont {Beran}}, \bibinfo {author} {\bibfnamefont
  {M.}~\bibnamefont {Zahradnik}}, \bibinfo {author} {\bibfnamefont
  {R.}~\bibnamefont {Antos}}, \bibinfo {author} {\bibfnamefont
  {M.}~\bibnamefont {Ku{\^c}era}},\ and\ \bibinfo {author} {\bibfnamefont
  {M.}~\bibnamefont {Veis}},\ }\bibfield  {title} {\bibinfo {title} {Optical
  and magneto-optical properties of bi substituted yttrium iron garnets
  prepared by metal organic decomposition},\ }\href
  {https://doi.org/10.1364/ome.6.001986} {\bibfield  {journal} {\bibinfo
  {journal} {Optical Materials Express}\ }\textbf {\bibinfo {volume} {6}},\
  \bibinfo {pages} {1986} (\bibinfo {year} {2016})}\BibitemShut {NoStop}%
\bibitem [{Sup(2023)}]{Supp}%
  \BibitemOpen
  \href@noop {} {\bibinfo {title} {Supplementary material that can be found in
  [enter url]}} (\bibinfo {year} {2023})\BibitemShut {NoStop}%
\bibitem [{\citenamefont {Sakoda}(2005)}]{Group-Theory-Phc}%
  \BibitemOpen
  \bibfield  {author} {\bibinfo {author} {\bibfnamefont {K.}~\bibnamefont
  {Sakoda}},\ }\href {https://doi.org/10.1007/b138376} {\emph {\bibinfo {title}
  {Optical Properties of Photonic Crystals}}}\ (\bibinfo  {publisher} {Springer
  Berlin, Heidelberg},\ \bibinfo {year} {2005})\BibitemShut {NoStop}%
\bibitem [{\citenamefont {{\'A}lvarez-Sanchis}\ \emph
  {et~al.}(2023)\citenamefont {{\'A}lvarez-Sanchis}, \citenamefont {Vidal},
  \citenamefont {Tretyakov},\ and\ \citenamefont
  {D{\'i}az-Rubio}}]{BIC-losses-destruction}%
  \BibitemOpen
  \bibfield  {author} {\bibinfo {author} {\bibfnamefont {J.~A.}\ \bibnamefont
  {{\'A}lvarez-Sanchis}}, \bibinfo {author} {\bibfnamefont {B.}~\bibnamefont
  {Vidal}}, \bibinfo {author} {\bibfnamefont {S.~A.}\ \bibnamefont
  {Tretyakov}},\ and\ \bibinfo {author} {\bibfnamefont {A.}~\bibnamefont
  {D{\'i}az-Rubio}},\ }\bibfield  {title} {\bibinfo {title} {Loss-induced
  performance limits of all-dielectric metasurfaces for terahertz sensing},\
  }\bibfield  {journal} {\bibinfo  {journal} {Physical Review Applied}\
  }\textbf {\bibinfo {volume} {19}},\ \href
  {https://doi.org/10.1103/PhysRevApplied.19.014009}
  {10.1103/PhysRevApplied.19.014009} (\bibinfo {year} {2023})\BibitemShut
  {NoStop}%
\bibitem [{\citenamefont {Kondratov}\ \emph {et~al.}(2016)\citenamefont
  {Kondratov}, \citenamefont {Gorkunov}, \citenamefont {Darinskii},
  \citenamefont {Gainutdinov}, \citenamefont {Rogov}, \citenamefont {Ezhov},\
  and\ \citenamefont {Artemov}}]{CMT-Chiral-Ports}%
  \BibitemOpen
  \bibfield  {author} {\bibinfo {author} {\bibfnamefont {A.~V.}\ \bibnamefont
  {Kondratov}}, \bibinfo {author} {\bibfnamefont {M.~V.}\ \bibnamefont
  {Gorkunov}}, \bibinfo {author} {\bibfnamefont {A.~N.}\ \bibnamefont
  {Darinskii}}, \bibinfo {author} {\bibfnamefont {R.~V.}\ \bibnamefont
  {Gainutdinov}}, \bibinfo {author} {\bibfnamefont {O.~Y.}\ \bibnamefont
  {Rogov}}, \bibinfo {author} {\bibfnamefont {A.~A.}\ \bibnamefont {Ezhov}},\
  and\ \bibinfo {author} {\bibfnamefont {V.~V.}\ \bibnamefont {Artemov}},\
  }\bibfield  {title} {\bibinfo {title} {Extreme optical chirality of plasmonic
  nanohole arrays due to chiral fano resonance},\ }\bibfield  {journal}
  {\bibinfo  {journal} {Physical Review B}\ }\textbf {\bibinfo {volume} {93}},\
  \href {https://doi.org/10.1103/PhysRevB.93.195418}
  {10.1103/PhysRevB.93.195418} (\bibinfo {year} {2016})\BibitemShut {NoStop}%
\bibitem [{\citenamefont {Caloz}\ and\ \citenamefont
  {Sihvola}(2019)}]{EM-chirality-Caloz}%
  \BibitemOpen
  \bibfield  {author} {\bibinfo {author} {\bibfnamefont {C.}~\bibnamefont
  {Caloz}}\ and\ \bibinfo {author} {\bibfnamefont {A.}~\bibnamefont
  {Sihvola}},\ }\href@noop {} {\bibinfo {title} {Electromagnetic chirality}}
  (\bibinfo {year} {2019}),\ \Eprint {https://arxiv.org/abs/1903.09087}
  {arXiv:1903.09087 [physics.optics]} \BibitemShut {NoStop}%
\bibitem [{\citenamefont {Fang}\ \emph {et~al.}(2011)\citenamefont {Fang},
  \citenamefont {Yu}, \citenamefont {Liu},\ and\ \citenamefont
  {Fan}}]{Optical-isolator}%
  \BibitemOpen
  \bibfield  {author} {\bibinfo {author} {\bibfnamefont {K.}~\bibnamefont
  {Fang}}, \bibinfo {author} {\bibfnamefont {Z.}~\bibnamefont {Yu}}, \bibinfo
  {author} {\bibfnamefont {V.}~\bibnamefont {Liu}},\ and\ \bibinfo {author}
  {\bibfnamefont {S.}~\bibnamefont {Fan}},\ }\bibfield  {title} {\bibinfo
  {title} {Ultracompact nonreciprocal optical isolator based on guided
  resonance in a magneto-optical photonic crystal slab},\ }\href
  {https://doi.org/10.1364/OL.36.004254} {\bibfield  {journal} {\bibinfo
  {journal} {Opt. Lett.}\ }\textbf {\bibinfo {volume} {36}},\ \bibinfo {pages}
  {4254} (\bibinfo {year} {2011})}\BibitemShut {NoStop}%
\bibitem [{\citenamefont {Braslavsky}(2007)}]{MCD-def}%
  \BibitemOpen
  \bibfield  {author} {\bibinfo {author} {\bibfnamefont {S.~E.}\ \bibnamefont
  {Braslavsky}},\ }\bibfield  {title} {\bibinfo {title} {Glossary of terms used
  in photochemistry, 3rd edition (iupac recommendations 2006)},\ }\href
  {https://doi.org/doi:10.1351/pac200779030293} {\bibfield  {journal} {\bibinfo
   {journal} {Pure and Applied Chemistry}\ }\textbf {\bibinfo {volume} {79}},\
  \bibinfo {pages} {293} (\bibinfo {year} {2007})}\BibitemShut {NoStop}%
\bibitem [{\citenamefont {Chen}\ \emph
  {et~al.}(2018{\natexlab{b}})\citenamefont {Chen}, \citenamefont {Kim},
  \citenamefont {Wong},\ and\ \citenamefont {Eleftheriades}}]{Huygens-review}%
  \BibitemOpen
  \bibfield  {author} {\bibinfo {author} {\bibfnamefont {M.}~\bibnamefont
  {Chen}}, \bibinfo {author} {\bibfnamefont {M.}~\bibnamefont {Kim}}, \bibinfo
  {author} {\bibfnamefont {A.~M.}\ \bibnamefont {Wong}},\ and\ \bibinfo
  {author} {\bibfnamefont {G.~V.}\ \bibnamefont {Eleftheriades}},\ }\bibfield
  {title} {\bibinfo {title} {Huygens' metasurfaces from microwaves to optics: a
  review},\ }\href {https://doi.org/doi:10.1515/nanoph-2017-0117} {\bibfield
  {journal} {\bibinfo  {journal} {Nanophotonics}\ }\textbf {\bibinfo {volume}
  {7}},\ \bibinfo {pages} {1207} (\bibinfo {year}
  {2018}{\natexlab{b}})}\BibitemShut {NoStop}%
\bibitem [{\citenamefont {Park}\ \emph {et~al.}(2021)\citenamefont {Park},
  \citenamefont {Asadchy}, \citenamefont {Zhao}, \citenamefont {Guo},
  \citenamefont {Wang},\ and\ \citenamefont
  {Fan}}]{CMT-Absorption-condition-1}%
  \BibitemOpen
  \bibfield  {author} {\bibinfo {author} {\bibfnamefont {Y.}~\bibnamefont
  {Park}}, \bibinfo {author} {\bibfnamefont {V.~S.}\ \bibnamefont {Asadchy}},
  \bibinfo {author} {\bibfnamefont {B.}~\bibnamefont {Zhao}}, \bibinfo {author}
  {\bibfnamefont {C.}~\bibnamefont {Guo}}, \bibinfo {author} {\bibfnamefont
  {J.}~\bibnamefont {Wang}},\ and\ \bibinfo {author} {\bibfnamefont
  {S.}~\bibnamefont {Fan}},\ }\bibfield  {title} {\bibinfo {title} {Violating
  kirchhoff's law of thermal radiation in semitransparent structures},\ }\href
  {https://doi.org/10.1021/acsphotonics.1c00612} {\bibfield  {journal}
  {\bibinfo  {journal} {ACS Photonics}\ }\textbf {\bibinfo {volume} {8}},\
  \bibinfo {pages} {2417} (\bibinfo {year} {2021})}\BibitemShut {NoStop}%
\bibitem [{\citenamefont {Asadchy}\ \emph {et~al.}(2015)\citenamefont
  {Asadchy}, \citenamefont {Faniayeu}, \citenamefont {Ra'di}, \citenamefont
  {Khakhomov}, \citenamefont {Semchenko},\ and\ \citenamefont
  {Tretyakov}}]{Broadband-Huygens}%
  \BibitemOpen
  \bibfield  {author} {\bibinfo {author} {\bibfnamefont {V.~S.}\ \bibnamefont
  {Asadchy}}, \bibinfo {author} {\bibfnamefont {I.~A.}\ \bibnamefont
  {Faniayeu}}, \bibinfo {author} {\bibfnamefont {Y.}~\bibnamefont {Ra'di}},
  \bibinfo {author} {\bibfnamefont {S.~A.}\ \bibnamefont {Khakhomov}}, \bibinfo
  {author} {\bibfnamefont {I.~V.}\ \bibnamefont {Semchenko}},\ and\ \bibinfo
  {author} {\bibfnamefont {S.~A.}\ \bibnamefont {Tretyakov}},\ }\bibfield
  {title} {\bibinfo {title} {Broadband reflectionless metasheets:
  Frequency-selective transmission and perfect absorption},\ }\href
  {https://doi.org/10.1103/PhysRevX.5.031005} {\bibfield  {journal} {\bibinfo
  {journal} {Phys. Rev. X}\ }\textbf {\bibinfo {volume} {5}},\ \bibinfo {pages}
  {031005} (\bibinfo {year} {2015})}\BibitemShut {NoStop}%
\end{thebibliography}%

\end{document}